\documentclass[twocolumn,english,aps,prl,superscriptaddress,10pt]{revtex4-1}

\usepackage{babel}
\usepackage{amsmath}
\usepackage{graphicx}
\usepackage{amssymb}

\makeatletter
\@ifundefined{textcolor}{}
{%
 \definecolor{BLACK}{gray}{0}
 \definecolor{WHITE}{gray}{1}
 \definecolor{RED}{rgb}{1,0,0}
 \definecolor{GREEN}{rgb}{0,1,0}
 \definecolor{BLUE}{rgb}{0,0,1}
 \definecolor{CYAN}{cmyk}{1,0,0,0}
 \definecolor{MAGENTA}{cmyk}{0,1,0,0}
 \definecolor{YELLOW}{cmyk}{0,0,1,0}
 }

\@ifundefined{definecolor}
 {\usepackage{color}}{}
\@ifundefined{definecolor}{\@ifundefined{definecolor}
 {\usepackage{color}}{}
}{}
\makeatother

\newcommand{\nio}{Na$_2$IrO$_3$}

\newcommand{\aio}{$A_2$IrO$_3$}
\newcommand{\rucl}{$\alpha$-RuCl$_3$}

\newcommand{\QQ}{\vec{Q}}

\newcommand{\hk}{Heisenberg-Kitaev}

\newcommand{\hc}{h_{{\rm c}0}}

\newcommand{\Ztwo}{\mathbb{Z}_2}
\newcommand{\Zthree}{\mathbb{Z}_3}

\graphicspath{{./}{./plots/}}

\begin{document}

\title{
Honeycomb-lattice Heisenberg-Kitaev model in a magnetic field:\\
Spin canting, metamagnetism, and vortex crystals
}

\author{Lukas Janssen}
\affiliation{Institut f\"ur Theoretische Physik, Technische Universit\"at Dresden,
01062 Dresden, Germany}
\author{Eric C.\ Andrade}
\affiliation{Instituto de F\'{i}sica de S\~ao Carlos, Universidade de S\~ao Paulo, C.P. 369,
S\~ao Carlos, SP,  13560-970, Brazil}
\author{Matthias Vojta}
\affiliation{Institut f\"ur Theoretische Physik, Technische Universit\"at Dresden,
01062 Dresden, Germany}


\begin{abstract}
The Heisenberg-Kitaev model is a paradigmatic model to describe the magnetism in honeycomb-lattice Mott insulators with strong spin-orbit coupling, such as $A_2$IrO$_3$ ($A = \mathrm{Na}, \mathrm{Li}$) and $\alpha$-RuCl$_3$.
Here we study in detail the physics of the Heisenberg-Kitaev model in an external magnetic field. Using a combination of Monte-Carlo simulations and spin-wave theory we map out the classical phase diagram for different directions of the magnetic field. Broken SU(2) spin symmetry renders the magnetization process rather complex, with sequences of phases and metamagnetic transitions. In particular, we find various large-unit-cell and multi-$Q$ phases including a vortex-crystal phase for a field in the $[111]$ direction. We also discuss quantum corrections in the high-field phase.
\end{abstract}



\maketitle


Magnets with strong spin-orbit coupling are currently in the focus of intense research, a primary motivation being the search for novel phases beyond the territory of the spin-isotropic Heisenberg model \cite{Jac09,Cha10,Nus15}. A paradigmatic example for nontrivial effects of spin-anisotropic interactions is Kitaev's honeycomb-lattice compass model, which realizes a $\Ztwo$ spin liquid of Majorana fermions \cite{Kit06}.

Materials with 4d and 5d transition-metal ions have been proposed to host strongly anisotropic exchange interactions between magnetic moments \cite{Jac09}. In particular, the insulating iridates \aio\ ($A = \mathrm{Na}, \mathrm{Li}$) are promising candidates to realize compass interactions \cite{Nus15}: Here, the Ir$^{4+}$ ions are arranged in a layered honeycomb-lattice structure, and the combined effect of spin-orbit coupling, Coulomb interaction, and exchange geometry generates $J_{{\rm eff}}=1/2$ moments subject to a combination of compass and Heisenberg interactions \cite{Shi09,Jac09}.
The resulting \hk\ (HK) model has been shown to host both spin-liquid and conventionally-ordered phases \cite{Cha10,Jia11,Reu11,Bha12,Cha13,perkins12,perkins13}. More recently, a similar proposal has been made for {\rucl} \cite{plumb14,sears15,baner16}.

Experimentally, these materials display magnetic order at low temperatures $T$, with {\nio} and {\rucl} realizing a collinear ``zigzag'' order \cite{Choi12,Ye12,sears15}. This state is indeed a ground state of the nearest-neighbor HK model, where it results from a competition of antiferromagnetic (AF) Kitaev and ferromagnetic (FM) Heisenberg interactions \cite{Cha13}. However, it has been pointed out that additional longer-range interactions of both Heisenberg and Kitaev type, as well as antisymmetric exchange, are likely to be present \cite{Kimchi11,rau14,perkins14,ioannis15,winter16}.
To date, the debate about the most appropriate effective spin model has not been settled for any of these materials, but there is growing experimental evidence for the pivotal role played by bond-dependent exchanges \cite{baner16,chun15}.
Hence, more theoretical predictions which can help to distinguish different models (see e.g. Refs.~\onlinecite{manni14,av14}) are called for.

\begin{figure*}
\includegraphics[width=\textwidth]{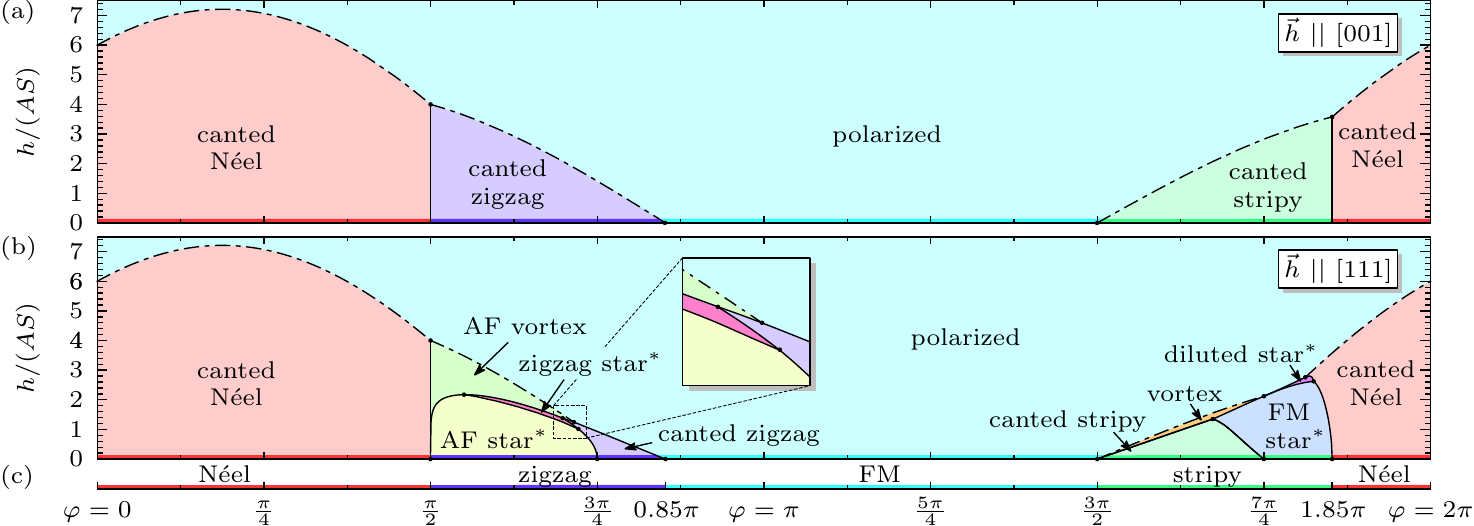}
\caption{
(a) Phase diagram of the classical HK model for $T\to0$, with $J=A \cos \varphi$, $K = A \sin \varphi$ and magnetic field $\vec h$ in the $[001]$ direction. Solid (dash-dotted) lines show first-order (second-order) phase transitions. For $\vec h$ along $[110]$ the phase diagram is identical. The dash-dotted line bounding the polarized phase represents $\hc(\varphi)$.
(b) Same as (a), but for $\vec h \parallel [1 1 1]$, showing a multitude of novel phases. Multi-$Q$ phases are asterisked; see text for details.
(c) Phase diagram for $h=0$ for comparison.
}
\label{fig:phase-diag}
\end{figure*}

In this paper, we study in detail the behavior of the HK model in an external magnetic field -- an issue of obvious experimental relevance which was covered only insufficiently in previous work \cite{Jia11, trousselet11, noteprevious, leekim13}. Using a combination of analytical and numerical techniques, we map out the low-$T$ phase diagram of the HK model in the classical limit for different field directions, Fig.~\ref{fig:phase-diag}. For a field applied along $[111]$ we find a surprisingly rich behavior -- very different from that of spin-symmetric Heisenberg models -- with a variety of canted phases, including multi-$Q$ and magnetic vortex states. The magnetization processes of both the zigzag and stripy phases are complex, with multiple metamagnetic transitions. We characterize all phases in detail and make concrete proposals to search for them in future experiments. We also compute the high-field magnetization for $S=1/2$ and quantify its deviation from saturation due to Kitaev interactions.


\paragraph{Model.}
We consider the spin-$S$ HK Hamiltonian \cite{Cha10,Cha13} on a honeycomb lattice in uniform magnetic field $\vec h$,
\begin{equation}
\mathcal{H}= J \sum_{\left\langle ij\right\rangle} \vec{S}_{i}\cdot\vec{S}_{j} + 2K \sum_{\left\langle ij\right\rangle
_{\gamma}} S_{i}^{\gamma}S_{j}^{\gamma}
- \vec h \cdot \sum_{i} \vec S_i.
\label{hk}
\end{equation}
Here $\gamma=x$, $y$, $z$ labels the three different links of the lattice.
The couplings may be parameterized as $J=A\cos\varphi$ and $K=A\sin\varphi$, where $A>0$ is an overall energy scale. The zero-field ground-state phase diagram of Eq.~\eqref{hk} for $S=1/2$ was first mapped out in Ref.~\onlinecite{Cha13}.
The classical-spin HK model, formally $S\to\infty$, reproduces \cite{perkins13} all phases of the spin-1/2 model except for the quantum spin liquid \cite{Cha13}, with $T\to0$
phase-boundary locations in reasonable agreement between quantum and classical models.
In this work we study the HK model at finite $\vec h$ and low $T$. We mainly focus on the large-$S$ limit, but also discuss quantum corrections to the high-field state.


%
\paragraph{Instability of the high-field state.}
We start with the high-field case, $h/S \gg A$. Taking the fully polarized state, with $\vec S_i \parallel \vec h$, as reference state, we have computed the magnon spectrum using spin-wave theory, for details see Sec.~I of Ref.~\onlinecite{suppl}.
In particular, magnons display a gap $\sim hS$ at large fields, and the vanishing of this gap as $h\searrow\hc$ usually signals the instability of the high-field state where magnon condensation drives the transition to a symmetry-broken canted state.

In the nonfrustrated cases we recover known results: For $\varphi = 0$ (AF Heisenberg model) the magnon energy vanishes at $\hc = 6 J S$ and wavevector $\QQ=\Gamma$. This is the ordering wavevector of the honeycomb-lattice N\'eel state, indicating a continuous transition at $\hc$ towards a canted deformation of the N\'eel phase. For $\varphi = \pi$ (FM Heisenberg model) there is no magnon softening at finite $h$, in accordance with the FM zero-field ground state.

The behavior significantly changes when a substantial Kitaev interaction is included, i.e., when the classical zero-field ground state is of zigzag or stripy form.
For a field in the $[001]$ or $[110]$ direction and $\pi/2 < \varphi < 0.85\pi$ (zigzag phase for $h=0$), a magnon mode becomes soft at the wavevector $\QQ=\mathrm M$. This corresponds to the ordering wavevector of the zigzag state, suggesting a direct transition between the high-field phase and a canted zigzag phase. By contrast, for $\vec h \parallel [111]$ the instability wavevector is $\QQ=\mathrm K$. Hence, as long as the instability is not preempted by a first-order transition above $\hc$, the phase below $\hc$ \emph{cannot} be smoothly connected to the zero-field zigzag ground state.
The same behavior is found above the stripy phase, $3\pi/2 < \varphi < 1.85\pi$: For $\vec h \parallel [001]$ or $[110]$ the high-field instability wavevector agrees with the zero-field ordering wavevector $\QQ=\mathrm M$, while for $\vec h \parallel [111]$ the instability wavevector is again $\QQ=\mathrm K$. Together, this points to the existence of novel intermediate phases when the field is in the diagonal direction.


\paragraph{Monte-Carlo simulations.}
To identify the low-temperature phases of the HK model at intermediate fields, we utilize classical Monte-Carlo (MC) simulations, combining single-site and parallel-tempering updates in order to equilibrate the spin configurations at low $T$, for details see Sec.~II of Ref.~\onlinecite{suppl}.
From the MC data, we compute the magnetization $\vec m$ and the static spin structure factor
$S_{\vec k} = N^{-1} \sum_{ij}\langle \vec S_i \cdot \vec S_j \rangle e^{i \vec k \cdot (\vec R_i - \vec R_j)}$,
where $\vec R_i$ is the lattice vector at site $i$ and $N$ the number of sites.

We perform simulations at various fixed values of $\varphi$, with $N \leq 2\cdot 24^2$ and in the temperature range $0.005 \lesssim T/\lvert JS^2 \rvert \lesssim 1.0$, to monitor the evolution of the magnetic state with increasing field. For selected parameter values we cool down the lowest-$T$ MC configuration to determine the corresponding classical ground state.
Our MC results reveal the existence of a number of nontrivial phases for $\vec h \parallel [111]$, to be described in detail below.

\begin{figure*}
\includegraphics[width=\textwidth]{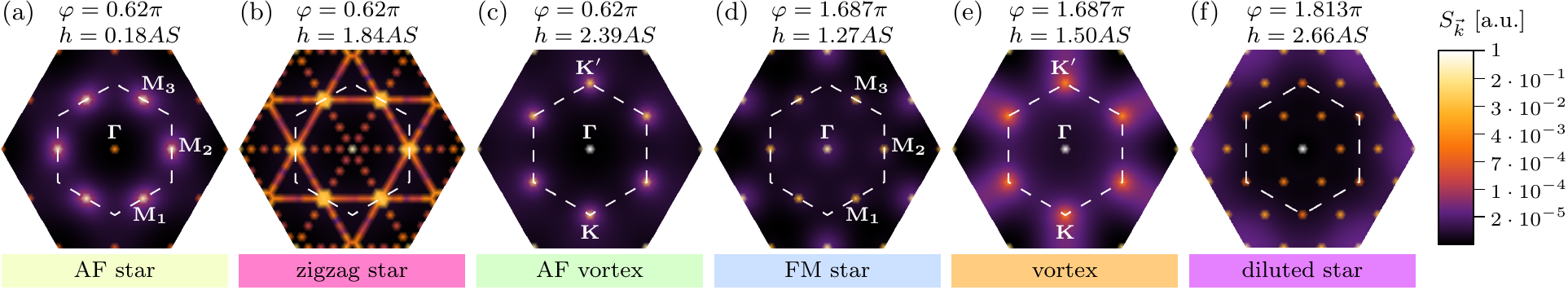}
\caption{Low-$T$ MC static structure factor for (a)--(c) $\varphi=0.62\pi$, (d), (e) $\varphi=1.687\pi$, and (f) $\varphi = 1.813\pi$, respectively, and different strengths $\vec h \parallel [1 1 1]$, showing magnetic Bragg peaks for various phases. Inner white dashed hexagon indicates the first Brillouin zone of the honeycomb lattice; high-symmetry points are labelled $\Gamma$, $\mathrm M_{1,2,3}$, and $\mathrm K, \mathrm K'$ in standard notation.
Here $N=2\cdot 18^2$ and $T/\lvert J S^2\rvert = 0.005$.
}
\label{fig:str-fct}
\end{figure*}

\begin{figure*}
\includegraphics[width=\textwidth]{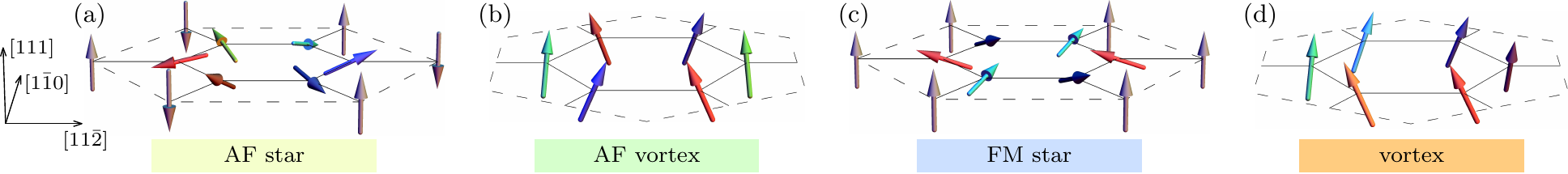}
\caption{Spin configurations of selected phases in three-dimensional spin space for $\vec h \parallel [111]$ (chosen vertical, see inset). The dashed lines denote the respective magnetic unit cell. $\varphi$ and $h$ are chosen as in Fig.~\ref{fig:str-fct}(a), (c), (d), and (e), respectively.
}
\label{fig:config}
\end{figure*}


\paragraph{Phases and phase diagram.}
In order to accurately compute the phase boundaries of the classical HK model in the $T\to0$ limit, we parameterize the spin configurations -- as deduced from the low-$T$ MC data -- of each of the phases analytically in terms of a set of angles (see Sec.~III of Ref.~\onlinecite{suppl}), and minimize the resulting energy by varying these angles at fixed $h$ and $\varphi$. We consider a total of 10 phases, and the comparison of their energies yields the $T\to0$ phase diagram as function of the coupling parameter $\varphi$ and the field strength $h$, Fig.~\ref{fig:phase-diag}.

For field $\vec h$ in the $[001]$ direction, Fig.~\ref{fig:phase-diag}(a), the field evolution is conventional: For infinitesimal field, the N\'eel, zigzag, and stripy ground states align perpendicular to the field (which reduces their degeneracy \cite{suppl}). With increasing $h$ they cant towards the field and undergo continuous transitions into the polarized phase at some $\hc$. This coincides with the instability field at which the high-field magnon becomes soft, thus validating our analysis. The same applies to $\vec h \parallel [110]$; we note that in both cases the field breaks the zero-field $\Zthree$ symmetry of combined $(2\pi/3)$-lattice and $(x\rightarrow y\rightarrow z)$-spin rotations.

For field $\vec h \parallel [111]$, Fig.~\ref{fig:phase-diag}(b), things are fundamentally different. Novel phases appear, with spin structure factors and real-space spin configurations illustrated in Figs.~\ref{fig:str-fct} and \ref{fig:config}, respectively.

We start the discussion with small fields.
Zigzag and stripy configurations cannot easily align perpendicular to the field direction, since these states have minimal energy only when pointing along one of the main cubic spin axes. As a result, canted zigzag (stripy) phases, with spins aligned along one of the cubic axes after projecting both in the plane perpendicular to $\vec h$, compete with other states. To understand these we note that the classical model has a continuous family of degenerate zero-field states which can be understood as different spin orientations of the dual AF (FM) Heisenberg model obtained from the Klein transformation \cite{Cha10, Cha13}, see Sec.~IV of Ref.~\onlinecite{suppl} for details.
While the zigzag (stripy) states are selected from this family at any $T>0$ by an order-from-disorder mechanism, the degeneracy is lifted even at $T=0$ by a finite field. For $\pi/2 < \varphi < 3\pi/4$ we find, on the one hand, that a small field favors an ``AF star'' pattern -- this is a triple-$Q$ state with $\QQ=\mathrm M$. It preserves the $\Zthree$ symmetry and has an 8-sublattice 8-site magnetic unit cell, where two sublattices are aligned in an alternating fashion along $[111]$, Fig.~\ref{fig:config}.
Similarly, for $7\pi/4 < \varphi \leq 1.85\pi$, the stripy zero-field ground state gives way to a triple-$Q$ ``FM star'' pattern with $\QQ=\mathrm M$ and a 4-sublattice 8-site unit cell.
On the other hand, single-$Q$ canted zigzag and stripy phases occur for $3\pi/4 \leq \varphi < 0.85\pi$ and $3\pi/2 < \varphi < 7\pi/4$, respectively. As in zero field, these break the $\Zthree$ symmetry.

Further nontrivial phases occur at elevated fields.
Above the stripy phase, for $7\pi/4 < \varphi < 1.85\pi$ a phase with 18-site magnetic unit cell is stabilized, with Bragg peaks at $\frac{2}{3}\mathrm{M}_i$ and $\mathrm K$. This multi-$Q$ phase, with $\Zthree$ preserved, can be understood as a ``diluted'' variant of the 8-site FM star configuration in which the 2 unit-cell spins that point in the $[111]$ direction are replaced by 12 spins which now are allowed to have small components perpendicular to $\vec h$.
In addition, for $3\pi/2 < \varphi < 7\pi/4$ a ``vortex'' phase, with $\Zthree$ preserved, is present at higher fields. It has a 6-site magnetic unit cell, with the spin components perpendicular to $\vec h$ winding around the center of each unit cell, describing a vortex-crystal configuration. In reciprocal space, this corresponds to single-$Q$ order with $\QQ=\mathrm K$; it is this order which emerges via magnon condensation from the high-field state at $\hc$.

\begin{figure*}
\includegraphics[width=\textwidth]{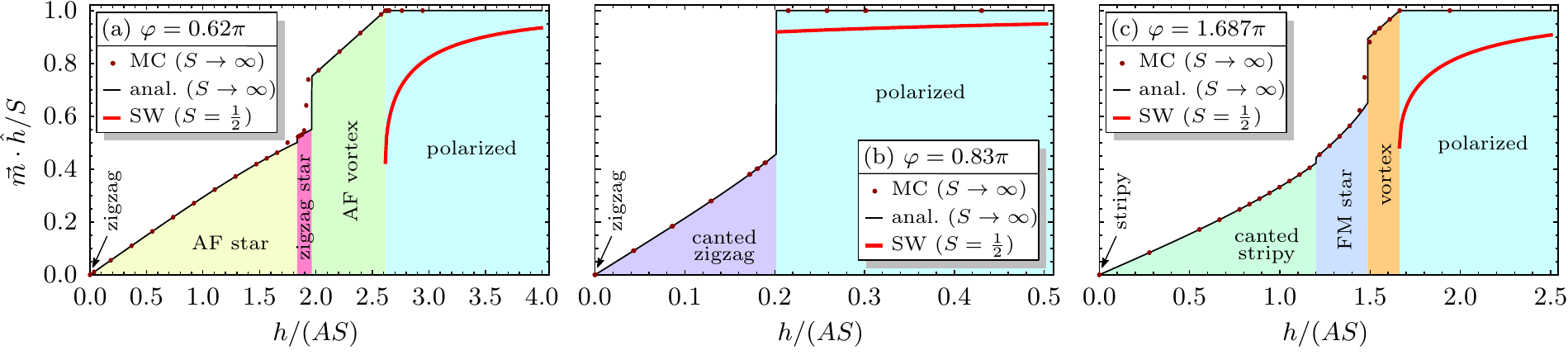}
\caption{Total magnetization of the HK model, $\vec m$, in field direction $\hat h \parallel [111]$ as function of field strength $h$ for different scans above the (a), (b) zigzag and (c) stripy states.
The curves show the classical magnetization obtained from cooling a low-$T$ MC configuration and analytical spin-pattern parametrization; in addition, the spin-wave result (SW) for $S=1/2$ is shown in the high-field phase, for details see text.
Color shading refers to phases in Fig.~\ref{fig:phase-diag}(b).}
\label{fig:magXh}
\end{figure*}

At elevated fields above the zigzag phase a ``zigzag star'' phase emerges for $0.55 \pi < \varphi < 0.72 \pi$. This state has an exceedingly large magnetic unit cell of 36 sites (or larger \cite{suppl}) with subdomains of zigzag and AF star patterns and a series of closely spaced Bragg peaks along lines in reciprocal space; it breaks $\Zthree$ and reflection symmetries.
Finally, for $\pi/2 < \varphi < 0.72\pi$ an ``AF vortex'' phase emerges. Similar to the vortex phase, this has a single-$Q$ modulation with $\QQ=\mathrm K$ and a 6-site magnetic unit cell, but now spins on different crystallographic sublattices wind around the hexagons' centers in opposite rotation directions.

Fig.~\ref{fig:str-fct} shows spin structure factors $S_{\vec k}$ of selected phases in a $[111]$ field. These have been obtained from low-$T$ MC simulations, implying that ``domain'' averaging over degenerate states with different but symmetry-related $Q$ has been performed,
such that single-$Q$ and multi-$Q$ phases with equivalent $Q$ cannot be distinguished. Single-domain $S_{\vec k}$ are displayed in Fig.~S6 of Ref.~\onlinecite{suppl}.


\paragraph{Magnetization process and phase transitions.}
The total magnetization, $\vec m = N^{-1} \sum_i \langle \vec S_i \rangle$, signals the various transitions.
For external field along $[001]$ or $[110]$, the $T\to 0$ magnetization in the classical limit is simply proportional to the field, $\vec m \propto \vec h$, for all canted phases before saturation at $\hc$. As already noted, the transition towards the polarized state is always continuous.

The situation again complexifies for $\vec h$ along $[111]$, where $\vec m$ is in general no longer parallel to the field axis. Fig.~\ref{fig:magXh} shows the total magnetization $\vec m$ projected onto the field direction $\hat h$ for different scans above the zigzag [(a), (b)] and stripy [(c)] phases, both from analytical parametrization of spin configurations as well as from cooled MC data. $\vec m \cdot \hat h$ increases -- in general nonlinearly -- with $h$ and exhibits characteristic jumps at first-order ``metamagnetic'' transitions below $\hc$. The transition towards the polarized phase at $\hc$ is in most cases continuous [Fig.~\ref{fig:magXh}(a) and (c)], and can hence be understood as condensation of a high-field magnon. Exceptions are $0.72\pi < \varphi < 0.85\pi$ and $\frac{7\pi}{4} < \varphi < 1.81\pi$, where the magnon instability as $h\searrow\hc$ is preempted by a first-order transition towards the canted zigzag phase and the diluted star phase, respectively [Fig.~\ref{fig:magXh}(b)].


\paragraph{Beyond the classical limit.}
So far, our analysis has been restricted to the classical limit $S\to\infty$. While this allowed us to construct the very rich phase diagram in Fig.~\ref{fig:phase-diag}, quantum fluctuations have a number of effects:
(i) The Kitaev spin-liquid phase appears in the vicinity of $J=0$ for $S=1/2$ \cite{Cha13}.
(ii) The classical phase boundaries get moderately shifted \cite{Jia11}. In particular,
the transition for $\pi/2<\varphi<3\pi/4$ ($7\pi/4<\varphi<1.85\pi$) from the zigzag (stripy) phase to the AF star (FM star) phase will be shifted from infinitesimal to finite field,
as the (canted) zigzag and stripy phases are stabilized by quantum fluctuations \cite{Cha13}.
However, the mere existence of the AF star and FM star phases in the quantum phase diagram is ensured by a hidden $\mathrm{SU}(2)$ symmetry, which forbids an order-from-disorder mechanism at the Klein points at $\varphi = 3\pi/4$ and $7\pi/4$, see Sec.~IV of Ref.~\onlinecite{suppl}.
(iii) Near $(\varphi,h) \simeq (0.71\pi,1.2AS)$, $(7\pi/4, 2.1AS)$, $(1.82\pi, 2.7 AS)$, when a large number of states are classically nearly degenerate, the classical phases may superseded by new quantum phases.
We therefore cannot exclude the possibility that the zigzag star and diluted star phases may be destroyed by quantum fluctuations. However, the vortex and AF vortex phases will survive as long as magnon interactions neither generate a fluctation-induced first-order transition nor shift the instability wavevector.
(iv) The magnetization of the high-field phase is no longer saturated unless $K=0$ or $h\to\infty$, as we discuss now.

We have used spin-wave theory to determine the magnetization in the high-field phase, with results for $S=1/2$ shown in Fig.~\ref{fig:magXh}, see Sec.~V of Ref.~\onlinecite{suppl} for details. We deduce a substantial reduction above both the stripy and zigzag phases; this reduction can be taken as a measure of deviation from SU(2) symmetry. As a result, for the experimentally relevant range $\varphi \sim (0.6\dots0.72) \pi$ the high-field phase is reached already at $m/m_{\rm sat} \approx 0.4 \dots 0.6$ for any field direction. For a continuous transition at $\hc$, the magnetization varies as $m- m(\hc) \propto (h-\hc)^\kappa$ with $\kappa=1/2$ in spin-wave (or mean-field) theory. Determining $\kappa$ exactly is subject of ongoing work.


\paragraph{Summary.}
We have shown that the classical HK model in a $[111]$ magnetic field displays a surprisingly rich phase diagram. In contrast to spin models with Heisenberg symmetry where collinear zero-field states typically turn into simple canted states, the HK model shows various complex large-unit-cell states. This demonstrates the potential of Kitaev interactions to produce vortex-crystal and other topologically nontrivial magnetic states \cite{daghofer16}.

Our findings call for new experiments on {\nio} and {\rucl} in a $[111]$ field~\cite{spinaxes}: (i) Neutron diffraction at intermediate fields in order to directly detect the novel phases identified here.
(ii) Inelastic neutron scattering at high fields to determine the wavevector where the high-field magnon becomes soft. This differing from the zero-field ordering wavevector indicates the existence of novel intermediate phases induced by Kitaev exchange. (iii) Careful low-$T$ magnetization measurements to search for possible metamagnetic transitions bounding such phases at low $T$.

Future theoretical work should investigate the quantum effects beyond linear-spin-wave theory, as well as models with longer-range and antisymmetric couplings in a magnetic field; these might yield even more complex field-induced phases.


\begin{acknowledgments}
We thank H.\ Jeschke, D.\ Joshi, N. Perkins, and S.\ Rachel for enlightening discussions. The computations were partially performed at the Center for Information Services and High Performance Computing (ZIH) at TU Dresden.
This research was supported by the DFG through SFB 1143 and GRK 1621
as well as by the Helmholtz association through VI-521.
E.\ C.\ A.\ was supported by FAPESP (Brazil) Grant No.\ 2013/00681-8.
\end{acknowledgments}



\begin{thebibliography}{99}
\bibitem{noteprevious}
Refs.~\onlinecite{Jia11, trousselet11} were restricted to small system sizes and omitted a detailed structure-factor analysis.
\end{thebibliography}
\end{document}


\title{
Supplemental material: \\
Honeycomb-lattice Heisenberg-Kitaev model in a magnetic field:\\
Spin canting, metamagnetism, and vortex crystals
}
\author{Lukas Janssen}
\affiliation{Institut f\"ur Theoretische Physik, Technische Universit\"at Dresden,
01062 Dresden, Germany}
\author{Eric C. Andrade}
\affiliation{Instituto de F\'{i}sica de S\~ao Carlos, Universidade de S\~ao Paulo, C.P. 369,
S\~ao Carlos, SP,  13560-970, Brazil}
\author{Matthias Vojta}
\affiliation{Institut f\"ur Theoretische Physik, Technische Universit\"at Dresden,
01062 Dresden, Germany}


%

\maketitle

\section{High-field instability from spin-wave theory}

In the asymptotic high-field limit all spins are aligned along the field axis. Magnon excitations are suppressed by a large energy gap. By lowering the field strength the magnon gap decreases and eventually vanishes at some critical field strength $\hc$. Below $\hc$ the high-field state becomes unstable, indicating a transition towards one of the various intermediate-field phases. (This is true as long as the continuous transition is not preempted by a first-order transition at some higher field $h_{\mathrm c} > \hc$.)
%
The wavevector at which the magnon gap closes then determines the ordering wavevector of the intermediate-field phase.

We parameterize the magnon excitations above the polarized ground state at high field by Holstein-Primakoff bosons $a_i$ and $b_i$ on the A and B sublattices of the honeycomb lattice.
It is convenient to use a spin-space frame obtained by rotating the cubic-axes basis $\vec e_x$, $\vec e_y$, $\vec e_z$ such that the magnetic field lies in the $3$-direction,
\begin{align} \label{eq:spin-frame}
 \vec e_1 & = \frac{(\vec e_z \times \vec h) \times \vec h}{\lvert (\vec e_z \times \vec h) \times \vec h \rvert},
 &
 \vec e_2 & = \frac{\vec e_z \times \vec h}{\lvert \vec e_z \times \vec h \rvert},
 &
 \vec e_3 & = \frac{\vec h}{\lvert \vec h \rvert}.
\end{align}
E.g., for field in the diagonal $[111]$ direction we choose the new spin-basis vectors
$\vec e_1 = (\vec e_x + \vec e_y - 2 \vec e_z)/\sqrt{6}$, $\vec e_2 = (-\vec e_x + \vec e_y)/\sqrt{2}$, and $\vec e_3 = (\vec e_x + \vec e_y + \vec e_z)/\sqrt{3}$.
To leading order in the $1/S$ expansion the spin operators in this basis read:
\begin{figure*}[!p]
\includegraphics[scale=0.66]{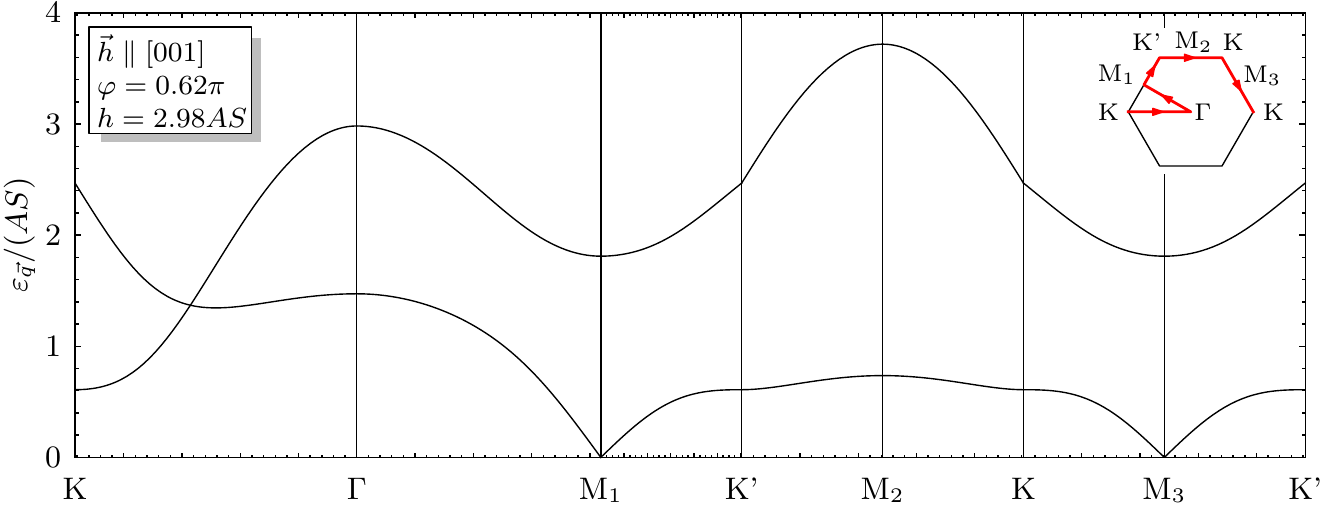}\hfill
\includegraphics[scale=0.66]{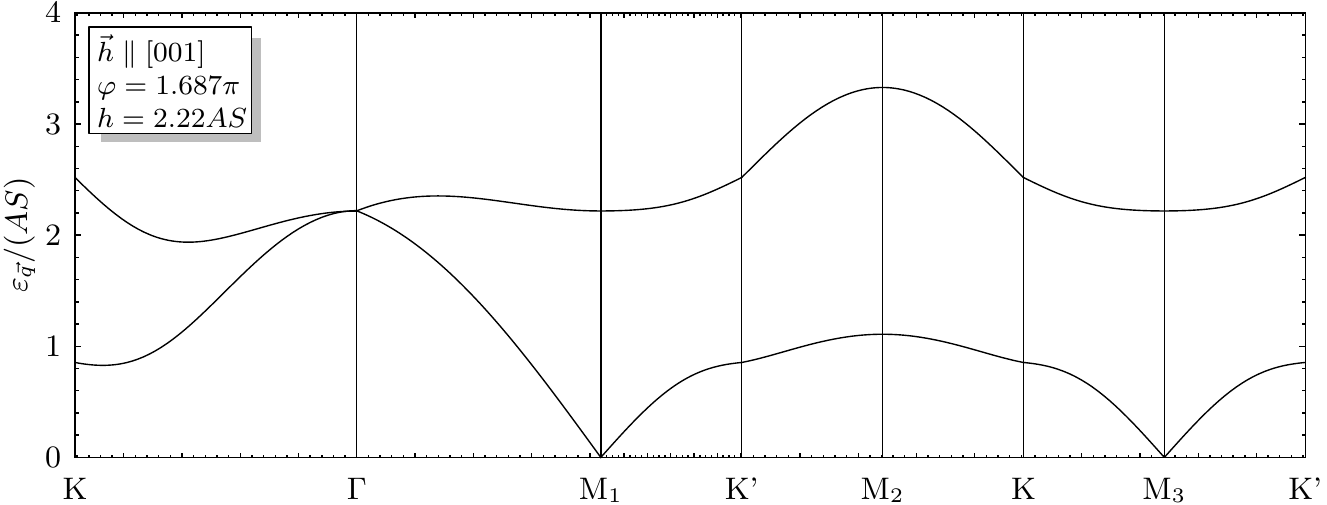}
\caption{Magnon excitation spectrum $\varepsilon_{\vec q}$ from linear spin-wave theory in the polarized phase for field in the $[001]$ direction and $h=\hc$ along high-symmetry lines in the Brillouin zone (see inset). The magnon gap vanishes at M$_1$ and M$_3$ for $\pi/2 < \varphi < \varphi_{\mathrm{c}2}$, when the transition is towards the canted zigzag phase (left panel), as well as for $3\pi/2 < \varphi < \varphi_{\mathrm{c}4}$, when the transition is towards the canted stripy phase (right panel).}
\label{fig:magnon-spectrum-001}
\end{figure*}
\begin{figure*}[!p]
\includegraphics[scale=0.65]{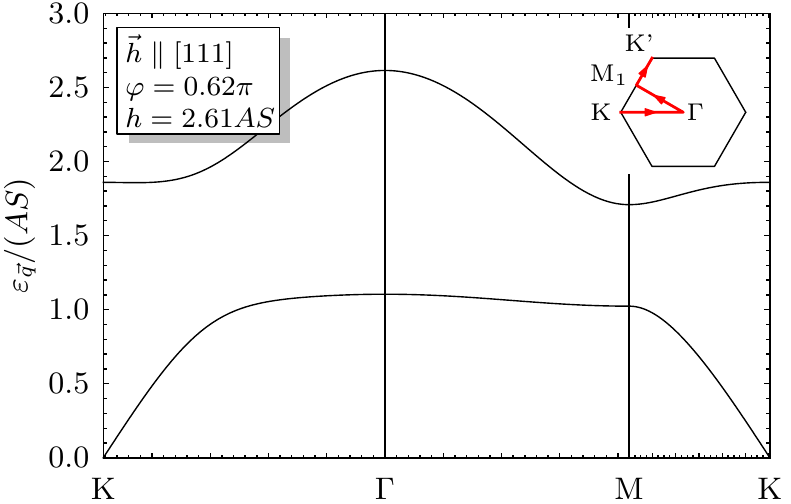}\hfill
\includegraphics[scale=0.65]{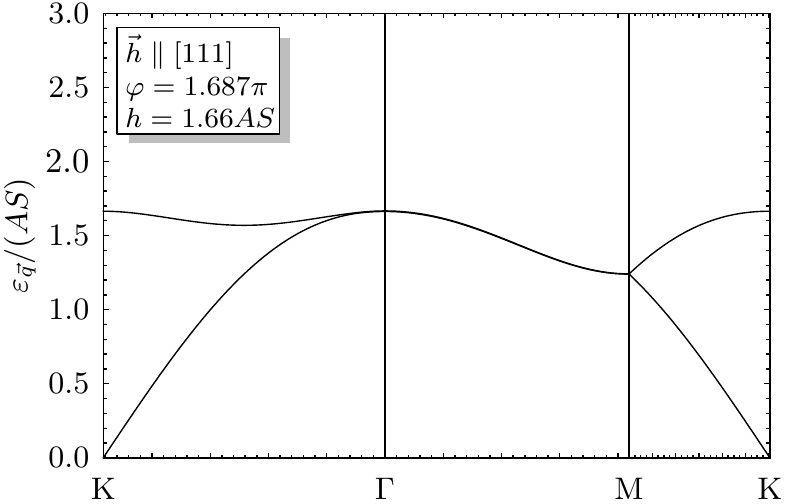}\hfill
\includegraphics[scale=0.65]{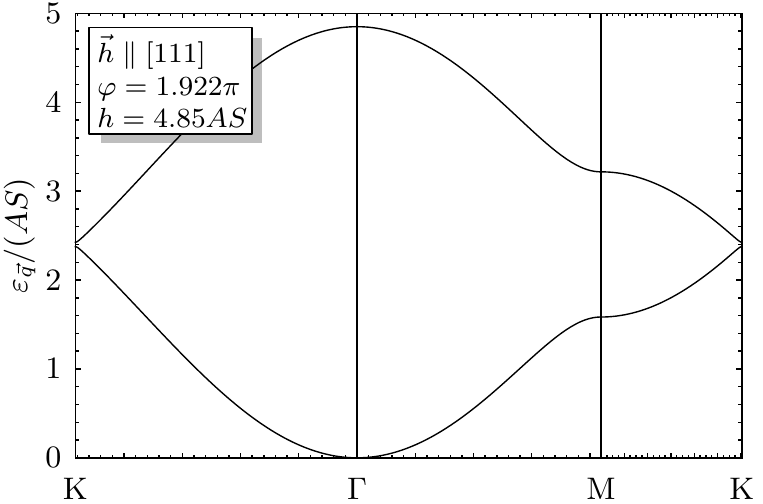}
\caption{Same as Fig.~\ref{fig:magnon-spectrum-001} for field in the $[111]$ direction. The magnon gap vanishes at the K points in the Brillouin zone for $\pi/2 < \varphi < \varphi_{\mathrm{c}1}$, when the transition is towards the AF vortex phase (left panel), as well as for $3\pi/2 < \varphi < 7\pi/4$, when the transition is towards the vortex phase (middle panel). For $\varphi_{\mathrm{c}3} < \varphi < 2\pi$ the instability wave vector is at the $\Gamma$ point, indicating the transition towards the canted N\'eel phase.
Note that due to the unbroken $\Zthree$ symmetry the magnon spectrum is the same at all three $\mathrm M$ points, in contrast to the situation when $\vec h \nparallel [111]$.
}
\label{fig:magnon-spectrum-111}
\end{figure*}
\begin{figure*}[!p]
\includegraphics[scale=0.65]{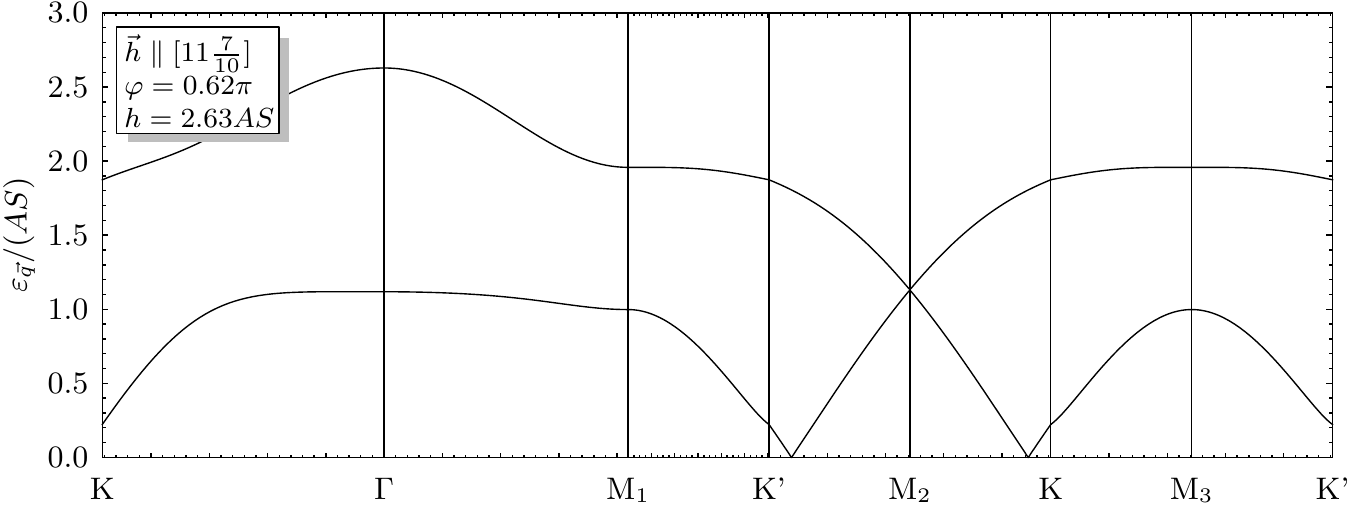}\hfill
\includegraphics[scale=0.65]{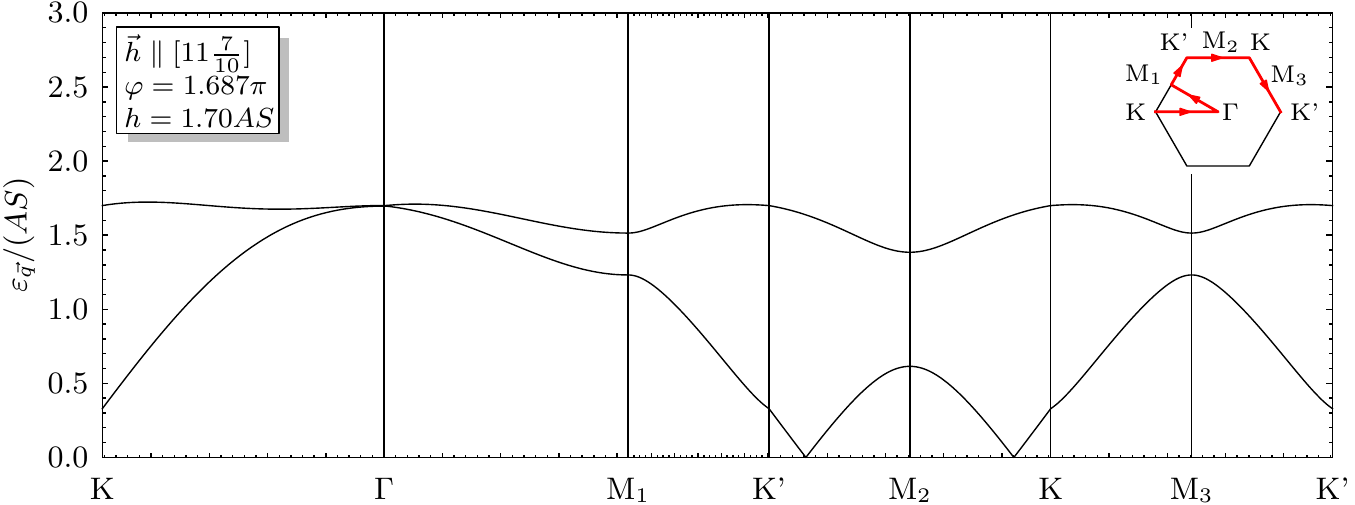}
\caption{Same as Fig.~\ref{fig:magnon-spectrum-001} for field in the $[11\frac{7}{10}]$ direction, when the magnon gap vanishes at incommensurate wavevectors between M$_2$ and K, K'.}
\label{fig:magnon-spectrum-1107}
\end{figure*}
\begin{table*}[!p]
\caption{Instability field strength $\hc$, instability wavevector $\QQ$, and intermediate-field phase below $\hc$ for different values of $\varphi$ and field directions. The transitions between the polarized and intermediate-field phases are always continuous except where indicated.
Here $\varphi_{\mathrm{c}1} \simeq 0.715\pi$, $\varphi_{\mathrm{c}2} = \pi - \arctan(1/2) \simeq 0.852\pi$, $\varphi_{\mathrm{c}3} \simeq 1.812\pi$, and $\varphi_{\mathrm{c}4} = 2\pi - \arctan(1/2) \simeq 1.852\pi$.
For $\vec h \parallel [111]$ the high-symmetry points $\mathrm M_1$, $\mathrm M_2$, and $\mathrm M_3$ have equivalent spectrum, but not for $\vec h \nparallel [111]$.}
\label{tab:instability-field}
\renewcommand{\arraystretch}{1.2}
\begin{tabular*}{\linewidth}{@{\extracolsep{\fill}} >{$}l<{$} | >{$}l<{$} >{$}l<{$} l | >{$}l<{$} >{$}l<{$} l | >{$}l<{$} >{$}l<{$} l}
\hline\hline
\multirow{2}{*}{$\varphi$} & \multicolumn{3}{c|}{$\vec h \parallel [001]$} & \multicolumn{3}{c|}{$\vec h \parallel [110]$} & \multicolumn{3}{c}{$\vec h \parallel [111]$}\\
	  & \hc(\varphi)/(AS) & \QQ & phase($h \nearrow \hc$) & \hc(\varphi)/(AS) & \QQ & phase($h \nearrow \hc$) & \hc(\varphi)/(AS) & \QQ & phase($h \nearrow \hc$) \\ \hline
0\dots \frac{\pi}{2} & 6 \cos\varphi + 4\sin\varphi & \Gamma & canted N\'eel
		     & 6 \cos\varphi + 4\sin\varphi & \Gamma & canted N\'eel
		     & 6 \cos\varphi + 4\sin\varphi & \Gamma & canted N\'eel \\
\frac{\pi}{2} \dots \varphi_{\mathrm{c}1}
    & 2 \cos\varphi + 4\sin\varphi & \mathrm{M}_1, \mathrm{M}_3 & canted zigzag
    & 2 \cos\varphi + 4\sin\varphi & \mathrm{M}_2 & canted zigzag
    & 3 \cos\varphi + 4\sin\varphi & \mathrm{K} & AF vortex \\
\varphi_{\mathrm{c}1} \dots \varphi_{\mathrm{c}2}
    & 2 \cos\varphi + 4\sin\varphi & \mathrm{M}_1, \mathrm{M}_3 & canted zigzag
    & 2 \cos\varphi + 4\sin\varphi & \mathrm{M}_2 & canted zigzag
    & \multicolumn{3}{c}{\it ---discontinuous $(h_\mathrm{c} > \hc)$---} \\
\varphi_{\mathrm{c}2} \dots \frac{3\pi}{2}
    & 0 & \Gamma & FM
    & 0 & \Gamma & FM
    & 0 & \Gamma & FM \\
\frac{3\pi}{2} \dots  \frac{7\pi}{4}
    & 4 \cos \varphi & \mathrm{M}_1, \mathrm{M}_3 & canted stripy
    & 4 \cos \varphi & \mathrm{M}_2 & canted stripy
    & 3 \cos \varphi & \mathrm{K} & vortex \\
\frac{7\pi}{4} \dots  \varphi_{\mathrm{c}3}
    & 4 \cos \varphi & \mathrm{M}_1, \mathrm{M}_3 & canted stripy
    & 4 \cos \varphi & \mathrm{M}_2 & canted stripy
    & \multicolumn{3}{c}{\it ---discontinuous $(h_\mathrm{c} > \hc)$---} \\
\varphi_{\mathrm{c}3} \dots \varphi_{\mathrm{c}4}
    & 4 \cos \varphi & \mathrm{M}_1, \mathrm{M}_3 & canted stripy
    & 4 \cos \varphi & \mathrm{M}_2 & canted stripy
    & 6 \cos \varphi + 4 \sin \varphi & \Gamma & canted N\'eel\\
\varphi_{\mathrm{c}4} \dots 2\pi
    & 6 \cos\varphi + 4\sin\varphi & \Gamma & canted N\'eel
    & 6 \cos\varphi + 4\sin\varphi & \Gamma & canted N\'eel
    & 6 \cos\varphi + 4\sin\varphi & \Gamma & canted N\'eel \\ \hline\hline
\end{tabular*}
\end{table*}
\begin{widetext}
\begin{align}
 \vec S_i & = (S - a^\dagger_i a_i) \vec e_3
 + \sqrt{\frac{S}{2}} (a_i + a^\dagger_i) \vec e_1
 + \ii \sqrt{\frac{S}{2}} (a_i - a^\dagger_i) \vec e_2 + \mathcal O(1/\sqrt{S}),
 \qquad i \in \mathrm A, \\
 \vec S_j & = (S - b^\dagger_j b_j) \vec e_3
 + \sqrt{\frac{S}{2}} (b_j + b^\dagger_j) \vec e_1
 + \ii \sqrt{\frac{S}{2}} (b_j - b^\dagger_j) \vec e_2 + \mathcal O(1/\sqrt{S}),
 \qquad j \in \mathrm B,
\end{align}
The spin-wave Hamiltonian in Fourier space then becomes (up to constant terms)
\begin{align}
 \mathcal H_\mathrm{SW} = S \sum_{\vec q \in \mathrm{BZ}} \biggl\{
 \left(\frac{h}{S}-3J-2K\right)
 \left(a^\dagger_{\vec q} a_{\vec q} + b^\dagger_{\vec q} b_{\vec q}\right)
 + \lambda_0(\vec q) a^\dagger_{\vec q} b_{\vec q}
 + \lambda_0^*(\vec q) b^\dagger_{\vec q} a_{\vec q}
 + \lambda_1(\vec q) a_{-\vec q} b_{\vec q} + \lambda_1^*(-\vec q) a^\dagger_{\vec q} b^\dagger_{-\vec q} \biggr\} + \mathcal{O}(1/S^0),
\end{align}
with
\begin{equation}
 \lambda_0(\vec q) =
 \begin{cases}
  (J + K) \left(\ee^{\ii \vec q \cdot \vec \delta_x} + \ee^{\ii \vec q \cdot \vec \delta_y}\right)
  + J \ee^{\ii \vec q \cdot \vec \delta_z}, & \quad \text{for } \vec h \parallel [001], \\
  \left(J + \frac{K}{2}\right) \left(\ee^{\ii \vec q \cdot \vec \delta_x} + \ee^{\ii \vec q \cdot \vec \delta_y}\right)
  + (J+K) \ee^{\ii \vec q \cdot \vec \delta_z}, & \quad \text{for } \vec h \parallel [110], \\
  \left(J+\frac{2}{3}K\right)\left(\ee^{\ii \vec q \cdot \vec \delta_x} + \ee^{\ii \vec q \cdot \vec \delta_y} + \ee^{\ii \vec q \cdot \vec \delta_z}\right), & \quad \text{for } \vec h \parallel [111],
 \end{cases}
\end{equation}
and
\begin{equation}
 \lambda_1(\vec q) =
 \begin{cases}
  K \left(\ee^{\ii \vec q \cdot \vec \delta_x} - \ee^{\ii \vec q \cdot \vec \delta_y} \right),
  & \quad \text{for } \vec h \parallel [001], \\
  \frac{K}{2} \left(-\ee^{\ii \vec q \cdot \vec \delta_x} - \ee^{\ii \vec q \cdot \vec \delta_y} + 2\ee^{\ii \vec q \cdot \vec \delta_z} \right),
  & \quad \text{for } \vec h \parallel [110], \\
  \frac{2}{3}K \left(
  \ee^{\ii \vec q \cdot \vec \delta_x - \frac{2\pi\ii}{3}}
  + \ee^{\ii \vec q \cdot \vec \delta_y + \frac{2\pi\ii}{3}}
  + \ee^{\ii \vec q \cdot \vec \delta_z}
  \right),
  & \quad \text{for } \vec h \parallel [111].
 \end{cases}
\end{equation}
\end{widetext}
Here, $\vec \delta_x$, $\vec \delta_y$, and $\vec \delta_z$ are the nearest-neighbor vectors on $x$, $y$, and $z$ bonds, respectively, of the honeycomb lattice. The leading-order piece of $\mathcal H_\mathrm{SW}$ is quadratic in boson operators and can be diagonalized analytically by means of a Bogoliubov transformation.

Resulting magnon spectra with $h \equiv \lvert \vec h \rvert$ tuned to the instability field strength $\hc$ are depicted for different coupling parameters $\varphi$ and field directions $\hat h \equiv \vec h / h$ in Figs.\ \ref{fig:magnon-spectrum-001}--\ref{fig:magnon-spectrum-1107}.
For large $h/S \gg |J|, |K|$ the minimum of the magnon dispersion is always at the $\Gamma$ point in the Brillouin zone.
As long as $|K| \ll |J|$ (i.e., when the zero-field ground state is a simple N\'eel or FM state), it remains at the $\Gamma$ point upon decreasing $h \searrow \hc$ at which it eventually vanishes; cf.\ right panel of Fig.~\ref{fig:magnon-spectrum-111}.
Above the zigzag and stripy zero-field ground states, however, the minimum of the dispersion shifts discontinuously from $\Gamma$ towards a finite wavevector as a function of field.
For field in the $[001]$ direction ($[110]$ direction) the instability wavevector at which the magnon gap eventually vanishes is at two (one) of the three inequivalent $\mathrm M$ points in the Brillouin zone, indicating a direct continuous transition towards the canted zigzag or stripy phase; see Fig.~\ref{fig:magnon-spectrum-001}. For field in the $[111]$ direction (left and middle panel of Fig.~\ref{fig:magnon-spectrum-111}), by contrast, the instability wavevector is at the $\mathrm K$ points, forbidding a direct continuous transition towards a simple canted deformation of the zero-field ground state. Remarkably, intermediate field directions lead to magnon softening at incommensurate wavevectors, see Fig.~\ref{fig:magnon-spectrum-1107} for field in the $[11 \frac{7}{10}]$ direction; we leave a detailed study of the resulting ordered states for future work.

We note that there is a linear band crossing point for field in the $[001]$ direction when $-0.148\pi < \varphi < 0.687\pi$. (An analogous band crossing point occurs when $0.852\pi < \varphi < 1.687\pi$.) This can be understood as a ``Dirac magnon'' that is located at the K point (and finite energy) in the Heisenberg limit $\varphi = 0$, and shifted from K towards the $\Gamma$ point (M$_2$ point) for finite $\varphi>0$ (finite $\varphi < 0$).\cite{magnonspectrum} Another such bosonic Dirac point is located at the opposite [with respect to the $\Gamma$ point (M$_2$ point) for $\varphi > 0$ ($\varphi < 0$)] wavevector. At $\varphi = 0.687\pi$ ($\varphi = -0.148\pi$) both merge and annihilate at the $\Gamma$ point (M$_2$ point).

Explicit values for the instability field strength $\hc$ and corresponding instability wavevectors are given in Table~\ref{tab:instability-field}. There, we have also indicated the special cases when the instability of the high-field magnon is preempted by a discontinuous transition, as obtained from the analytical parameterization of phases (Sec.~\ref{sec:parametrization}). In all other cases, the transition from polarized towards intermediate-field phases is continuous, and we have checked that the instability field strength $\hc$ indeed then always coincides with the critical field strength $h_\mathrm{c}$ as obtained from the parametrization (Sec.~\ref{sec:parametrization}), as well as with $h_\mathrm{c}$ from the MC data (Sec.~\ref{sec:mc}).
This also serves as an independent verification of the numerics.

We also note that the magnetization process in the HK model on the 3D hyperhoneycomb lattice in $[111]$ field appears to be similarly complex as found here. 
This is because the magnon instability at $\hc$ (which happens to coincide with $\hc$ for the 2D honeycomb lattice) occurs \emph{above} the metamagnetic first-order transitions found in the MC simulations. 
This has apparently been overlooked in the previous analysis.\cite{lee2014}


\section{Monte-Carlo simulations}\label{sec:mc}

To identify the intermediate-field phases, we study the large-$S$ limit of the HK model by employing a combination of classical Monte Carlo (MC) simulations and energy minimization. We work on honeycomb lattices of size $L\times L$ with periodic boundary conditions.
The lattices are spanned by the primitive lattice vectors $\vec{a}_{1\left(2\right)}=\left(3/2,\pm\sqrt{3}/2\right)$,
with each unit cell containing two sites amounting to a total number of spins of $N=2L^{2}$.

We perform equilibrium MC simulations using single-site updates with a combination of the
heat-bath and microcanonical (or over-relaxation) algorithms,\cite{berg_mc} with typically
$10^{7}$ MC steps per spin. We combine these updates with the parallel-tempering algorithm \cite{partemp_jpsj}
in order to efficiently equilibrate the MC configurations at very low $T$. From the MC data, we compute the uniform magnetization in the field direction (Fig.~\figmagXh\ in the main text)
\begin{equation}\label{eq:mag}
\frac{\vec m \cdot \hat h}{S}=\left \langle \frac{1}{N}\sum_{i}\frac{\vec{S}_{i}}{S}\cdot\frac{\vec h}{\lvert \vec h \rvert}\right\rangle,
\end{equation}
where $\langle \cdots \rangle $ denotes MC average, as well as the static spin structure factor (Fig.~\figstrfct\ in the main text)
\begin{equation}\label{eq:sfactor}
S_{\vec{k}} = \left\langle \vec{S}(\vec{k})\cdot \vec{S}(-\vec{k})\right\rangle,
\end{equation}
where
\begin{equation}\label{eq:sfourier}
\vec{S}(\vec{k}) = \frac{1}{\sqrt{N}}\sum_{i}\vec{S}_{i}e^{-i\vec{k}\cdot\vec{R}_{i}},
\end{equation}
is the Fourier transform of a given spin configuration and $\vec R_i$ is the lattice vector at site $i$.

To find the classical ground state, we start from a MC spin configuration obtained at low $T$ (typically $T/\lvert J S^2 \rvert \sim 0.005$) and then iteratively align the spins with their local fields\cite{hloc} $\vec{h}_{i}^\text{loc}$,
\begin{equation}\label{eq:min}
\vec{S}_{i} = \frac{\vec{h}_{i}^\text{loc}}{\left|\vec{h}_{i}^\text{loc}\right|} S.
\end{equation}
Convergence is reached after the largest update in a lattice sweep, $|\vec{S}_{i}^\text{new}-\vec{S}_{i}^\text{old}|_\text{max}/S$, is smaller than $10^{-12}$.
Because of the several competing ground states, it is important to start from unbiased MC configurations in order to obtain the
correct classical ground state.

\begin{figure*}[btp]
\includegraphics[width=\textwidth]{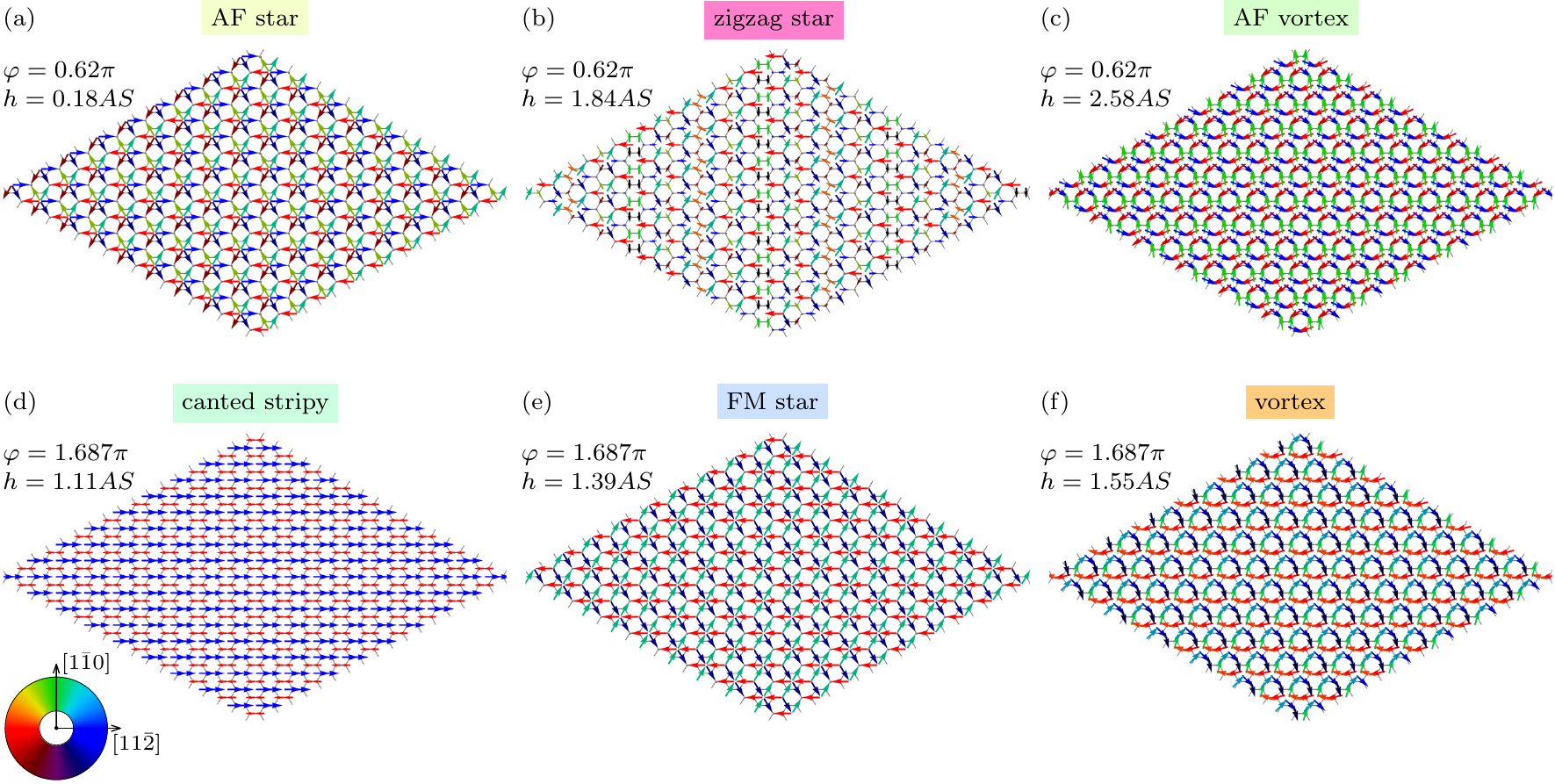}
\caption{Real-space spin configuration snapshots from cooled MC data, projected onto the plane perpendicular to $\vec h \parallel [111]$, for $N=2 \cdot 18^2$ spins.
The colors of the arrows refer to the spin directions in this perpendicular plane as a code to guide the eye (see inset); the arrows' lengths refer to the relative spin magnitudes in this plane (cf.\ the spin projections in Fig.~\ref{fig:spin-configs}).
}
\label{fig:mc-snapshots}
\end{figure*}

We performed extensive field scans at $\varphi = 0.57\pi$, $0.62\pi$, $0.733\pi$, $0.83\pi$, $1.578\pi$, $1.687\pi$, $1.813\pi$, and $1.922\pi$ for the $[111]$ field direction, $\varphi = 0.62\pi$ and $1.687\pi$ for the $[001]$ direction, as well as $\varphi = 1.687\pi$ for the $[110]$ field direction, with system sizes up to $L=24$. For field in the $[111]$ direction we find a total of 10 phases at finite $h$. Spin-configuration snapshots from cooled MC data are depicted for selected parameter values in Fig.~\ref{fig:mc-snapshots}.


\section{Parametrization of phases}\label{sec:parametrization}

The low-$T$ spin configurations obtained from the MC simulations allows the deduction of the symmetries, unit-cell sizes,\cite{largecell} and sublattice structure of the different phases. We make use of this information by parametrizing the spin configurations in terms of a set of angles, which then are optimized at fixed model parameters $\varphi$ and $\vec h$ to determine the state of lowest energy. Doing this for all phases enables a comparison of energies from which we deduce the classical phase diagram in the low-temperature limit.

\begin{table*}[p!]
\caption{Ans\"atze for angles $\phi_i$ and $\theta_i$ in Eq.~\eqref{eq:spin-ansatz} for parametrization of spin $\vec S_i$ at lattice site $\vec R_i$ within one magnetic unit cell. Here, $\vec R_i$ is measured in units of the lattice constant from the center of the first hexagon.}
\label{tab:angles}
\begin{minipage}[t]{\columnwidth}
\begingroup
\renewcommand{\arraystretch}{1.077}
\mbox{}\\[-\baselineskip]
\begin{tabular*}{\columnwidth}{@{\extracolsep{\fill}} lclll}
\hline\hline
Phase & $i$ & $\vec R_i$ & $\phi_i$ & $\theta_i$ \\ \hline
polarized 	& 1 & $(1,0)$ & 0 & 0 \\
		& 2 & $(\cos \frac{\pi}{3},\sin \frac{\pi}{3})$ & 0 & 0 \\
canted zigzag	& 1 & $(1,0)$ & $\pi$ & $\theta$ \\
		& 2 & $(\cos \frac{\pi}{3},\sin \frac{\pi}{3})$ & $\pi$ & $\theta$ \\
		& 3 & $(\cos \frac{2\pi}{3},\sin \frac{2\pi}{3})$ & $0$ & $\theta'$ \\
		& 4 & $(-1,0)$ & $0$ & $\theta'$ \\
AF star		& 1 & $(1,0)$ & $0$ & $\theta$ \\
		& 2 & $(\cos \frac{\pi}{3},\sin \frac{\pi}{3})$ & $\frac{\pi}{3}$ & $\theta'$ \\
		& 3 & $(\cos \frac{2\pi}{3},\sin \frac{2\pi}{3})$ & $\frac{2\pi}{3}$ & $\theta$ \\
		& 4 & $(-1,0)$ & $\pi$ & $\theta'$ \\
		& 5 & $(\cos \frac{4\pi}{3},\sin \frac{4\pi}{3})$ & $\frac{4\pi}{3}$ & $\theta$ \\
		& 6 & $(\cos \frac{5\pi}{3},\sin \frac{5\pi}{3})$ & $\frac{5\pi}{3}$ & $\theta'$ \\
		& 7 & $(2,0)$ & $0$ & $\pi$ \\
		& 8 & $(2\cos \frac{\pi}{3}, 2\sin \frac{\pi}{3})$ & $0$ & $0$ \\
AF vortex	& 1 & $(1,0)$ & $\frac{2\pi}{3} - \delta$ & $\theta$ \\
		& 2 & $(\cos \frac{\pi}{3},\sin \frac{\pi}{3})$ & $\frac{5\pi}{3} + \delta$ & $\theta$ \\
		& 3 & $(\cos \frac{2\pi}{3},\sin \frac{2\pi}{3})$ & $\frac{4\pi}{3}-\delta$ & $\theta$ \\
		& 4 & $(-1,0)$ & $\frac{\pi}{3}+\delta$ & $\theta$ \\
		& 5 & $(\cos \frac{4\pi}{3},\sin \frac{4\pi}{3})$ & $-\delta$ & $\theta$ \\
		& 6 & $(\cos \frac{5\pi}{3},\sin \frac{5\pi}{3})$ & $\delta$ & $\theta$ \\
vortex		& 1 & $(1,0)$ & $\frac{5\pi}{3} - \delta$ & $\theta$ \\
		& 2 & $(\cos \frac{\pi}{3},\sin \frac{\pi}{3})$ & $\frac{5\pi}{3} + \delta$ & $\theta$ \\
		& 3 & $(\cos \frac{2\pi}{3},\sin \frac{2\pi}{3})$ & $\frac{\pi}{3}-\delta$ & $\theta$ \\
		& 4 & $(-1,0)$ & $\frac{\pi}{3}+\delta$ & $\theta$ \\
		& 5 & $(\cos \frac{4\pi}{3},\sin \frac{4\pi}{3})$ & $\pi-\delta$ & $\theta$ \\
		& 6 & $(\cos \frac{5\pi}{3},\sin \frac{5\pi}{3})$ & $\pi+\delta$ & $\theta$ \\
zigzag star	& $1, \dots, 9$ & $(\frac{3i-1}{2},\frac{\sqrt{3}(i-1)}{2})$ & $\phi_i$ & $\theta_i$ \\
		& $10, \dots, 18$ & $(\frac{3i-29}{2},\frac{\sqrt{3}(i-9)}{2})$ & $\phi_i$ & $\theta_i$ \\
		& $19, \dots, 27$ & $(\frac{3i-55}{2},\frac{\sqrt{3}(i-17)}{2})$ & $\phi_i$ & $\theta_i$ \\
		& $28, \dots, 36$ & $(\frac{3i-83}{2},\frac{\sqrt{3}(i-25)}{2})$ & $\phi_i$ & $\theta_i$ \\ \hline\hline
\end{tabular*}\endgroup
\end{minipage}\hfill
\begin{minipage}[t]{\columnwidth}
\mbox{}\\[-\baselineskip]
\begin{tabular*}{\columnwidth}{@{\extracolsep{\fill}} lclll}
\hline\hline
Phase & $i$ & $\vec R_i$ & $\phi_i$ & $\theta_i$ \\ \hline
canted N\'eel 	& 1 & $(1,0)$ & 0 & $\theta$ \\
		& 2 & $(\cos \frac{\pi}{3},\sin \frac{\pi}{3})$ & $\pi$ & $\theta$ \\
canted stripy	& 1 & $(1,0)$ & $\pi$ & $\theta$ \\
		& 2 & $(\cos \frac{\pi}{3},\sin \frac{\pi}{3})$ & $0$ & $\theta'$ \\
		& 3 & $(\cos \frac{2\pi}{3},\sin \frac{2\pi}{3})$ & $0$ & $\theta'$ \\
		& 4 & $(-1,0)$ & $\pi$ & $\theta$ \\
FM star		& 1 & $(1,0)$ & $\pi$ & $\theta$ \\
		& 2 & $(\cos \frac{\pi}{3},\sin \frac{\pi}{3})$ & $\frac{\pi}{3}$ & $\theta$ \\
		& 3 & $(\cos \frac{2\pi}{3},\sin \frac{2\pi}{3})$ & $\frac{5\pi}{3}$ & $\theta$ \\
		& 4 & $(-1,0)$ & $\pi$ & $\theta$ \\
		& 5 & $(\cos \frac{4\pi}{3},\sin \frac{4\pi}{3})$ & $\frac{\pi}{3}$ & $\theta$ \\
		& 6 & $(\cos \frac{5\pi}{3},\sin \frac{5\pi}{3})$ & $\frac{5\pi}{3}$ & $\theta$ \\
		& 7 & $(2,0)$ & $0$ & $0$ \\
		& 8 & $(2\cos \frac{\pi}{3}, 2\sin \frac{\pi}{3})$ & $0$ & $0$ \\
diluted star	& 1 & $(1,0)$ & $\pi$ & $\theta$ \\
		& 2 & $(\cos \frac{\pi}{3},\sin \frac{\pi}{3})$ & $\frac{\pi}{3}$ & $\theta$ \\
		& 3 & $(\cos \frac{2\pi}{3},\sin \frac{2\pi}{3})$ & $\frac{5\pi}{3}$ & $\theta$ \\
		& 4 & $(-1,0)$ & $\pi$ & $\theta$ \\
		& 5 & $(\cos \frac{4\pi}{3},\sin \frac{4\pi}{3})$ & $\frac{\pi}{3}$ & $\theta$ \\
		& 6 & $(\cos \frac{5\pi}{3},\sin \frac{5\pi}{3})$ & $\frac{5\pi}{3}$ & $\theta$ \\
		& 7  & $(2,0)$ & $0$ & $\theta$ \\
		& 8  & $(2\cos \frac{\pi}{3}, 2\sin \frac{\pi}{3})$ & $\frac{4\pi}{3}$ & $\theta$ \\
		& 9  & $(2\cos \frac{2\pi}{3}, 2\sin \frac{2\pi}{3})$ & $\frac{2\pi}{3}$ & $\theta$ \\
		& 10 & $(-2,0)$ & $0$ & $\theta$ \\
		& 11 & $(2\cos \frac{4\pi}{3}, 2\sin \frac{4\pi}{3})$ & $\frac{4\pi}{3}$ & $\theta$ \\
		& 12 & $(2\cos \frac{5\pi}{3}, 2\sin \frac{5\pi}{3})$ & $\frac{2\pi}{3}$ & $\theta$ \\
		& 13 & $(2,-\sqrt{3})$ & $\frac{4\pi}{3}$ & $\theta'$\\
		& 14 & $(\frac{5}{2},-\frac{\sqrt{3}}{2})$ & $\frac{4\pi}{3}$ & $\theta'$ \\
		& 15 & $(\frac{5}{2},\frac{\sqrt{3}}{2})$ & $\frac{2\pi}{3}$ & $\theta'$ \\
		& 16 & $(2,\sqrt{3})$ & $\frac{2\pi}{3}$ & $\theta'$ \\
		& 17 & $(\frac{1}{2},\frac{3\sqrt{3}}{2})$ & $0$ & $\theta'$ \\
		& 18 & $(-\frac{1}{2},\frac{3\sqrt{3}}{2})$ & $0$ & $\theta'$ \\ \hline\hline
\end{tabular*}
\end{minipage}
\end{table*}

Using the rotated basis as defined in Eq.~\eqref{eq:spin-frame}, the spin $\vec S_i$ at site $i$ can be parametrized as
\begin{equation} \label{eq:spin-ansatz}
 \vec S_i = S \left(
 \vec e_1 \sin \theta_i \cos \phi_i +
 \vec e_2 \sin \theta_i \sin \phi_i +
 \vec e_3 \cos \theta_i
 \right).
\end{equation}
In the polarized phase we have $\theta_i \equiv 0$, while $\theta_i > 0$ defines a canted state.
For given coupling parameter $\varphi$ and magnetic field $\vec h = h \vec e_3$ our ans\"atze for the angles $\theta_i$ and $\phi_i$ as obtained from the MC simulations are given in Table \ref{tab:angles}. Except for the vortex, AF vortex, and zigzag star phases the spin projections onto the plane perpendicular to $\vec h$ ($\vec e_1$-$\vec e_2$ plane) are locked on the directions of the cubic-axes projections $\vec e_3 \times (\vec e_x \times \vec e_3)$, $\vec e_3 \times (\vec e_y \times \vec e_3)$, and $\vec e_3 \times (\vec e_z \times \vec e_3)$, see Fig.~\ref{fig:mc-snapshots}. For these phases we therefore have $\phi_i \in \{0, \frac{\pi}{3}, \frac{2\pi}{3}, \pi, \frac{4\pi}{3}, \frac{5\pi}{3}\}$, and we may minimize with respect to the field-dependent canting angles $\theta_i$ only. In each case we in fact find that there are at most only two different possible $\theta$ angles [indicated by the at most two different lengths of the spin projections in Fig.~\ref{fig:mc-snapshots}(a), (d), and (e)]. This makes the computation of the minimized energy of a given classical state and their comparison among different states numerically cheap. In the cases of the vortex and AF vortex phases the $\phi$ angles are not locked onto the projection of the cubic-axes direction. However, we find that the classical energy in these cases in fact becomes independent of the angle $\delta$ that determines the (uniform) deviation from the cubic-axes locking. (The MC data show that thermal fluctuations lift this degeneracy by an order-from-disorder mechanism.) By contrast, for the zigzag star phase we do not use any particular ansatz for the configuration, except for the fact (again as obtained from the cooled MC data) that the magnetic unit cell spans $2 \times 9$ crystallographic unit cells.\cite{largecell} The explicit assumptions for $\theta_i$ and $\phi_i$ for all states are summarized in Table~\ref{tab:angles}.

\begin{figure*}[!p]
\includegraphics[width=\textwidth]{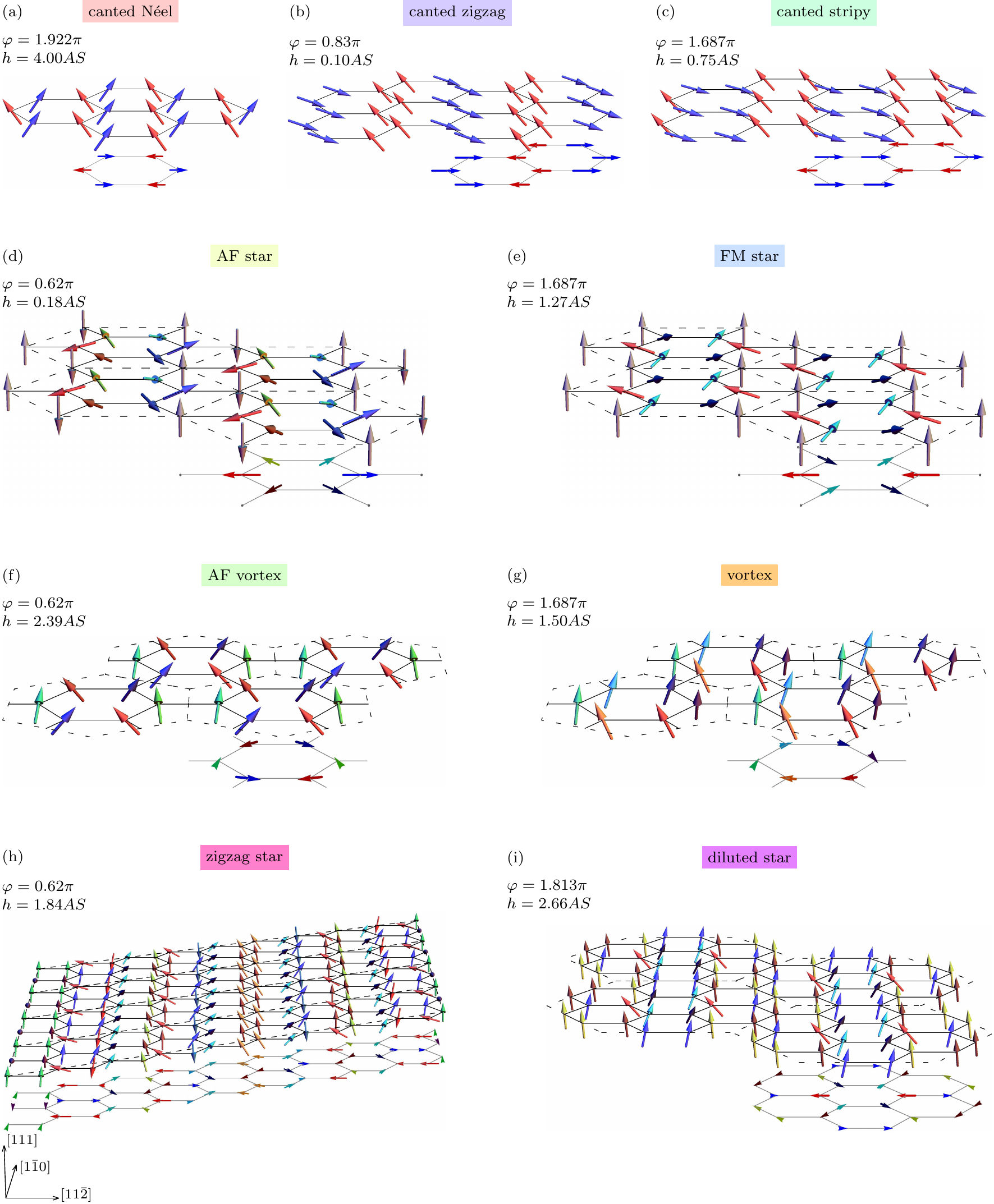}
\caption{Real-space spin configurations from analytical parametrization for various states. Here we have aligned the $[111]$ spin-space axis perpendicular to the real-space lattice.
Dashed: magnetic unit cell.
Gray lattices: spin projections onto the plane perpendicular to $\vec h$.
The colors of the arrows refer to the spin directions in this perpendicular plane (see inset of Fig.~\ref{fig:mc-snapshots}).
}
\label{fig:spin-configs}
\end{figure*}
\begin{figure*}[!p]
\includegraphics[width=0.45\linewidth]{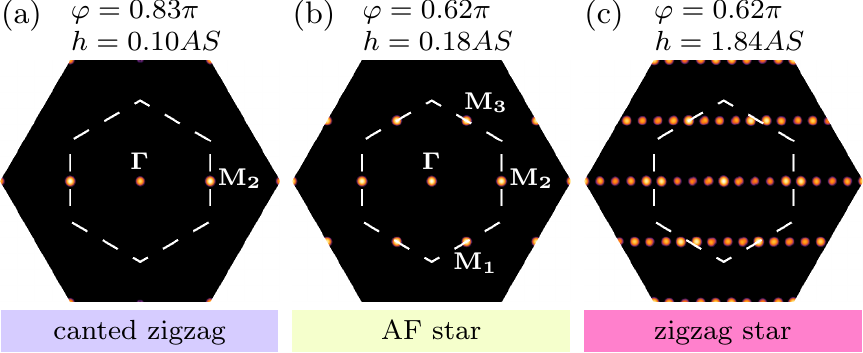}
\caption{
Single-domain static spin structure factor from analytical parametrization for different values of $\varphi$ and $\vec h \parallel [111]$, allowing us to distinguish between (a) the single-$Q$ order of the canted zigzag phase and (b) the triple-$Q$ AF star pattern. The zigzag star phase (c) has a total of 18 inequivalent Bragg peaks within the first Brillouin zone. Averaging over the six symmetry-related ground states, obtained by $2\pi/3$ rotation and inversion, yields the hexagram pattern as seen in the MC spin structure, Fig.~{\figstrfct}(b) in the main text.
The inner white dashed hexagon indicates the location of the first Brillouin zone of the honeycomb lattice. In all cases, the signal at $\Gamma$ arises from the uniform magnetization component in field direction.
For visualization purposes, we have replaced the $\delta$ peaks of the infinite-size system by finite-width Gaussian distributions.
}
\label{fig:strctr-fctr_ana}
\end{figure*}

The parametrization allows the straightforward comparison of the minimized energies of the various states and the deduction of the phase boundaries for arbitrary coupling parameter $\varphi$ and field strength $h$ under the assumption that no further states (not parametrized in Table~\ref{tab:angles}, and missed by the MC scans) are stabilized somewhere in the phase diagram. The result is depicted in Fig.~\figpd\ in the main text. In addition to the quadruple point at $(\varphi, h/(AS)) = (7\pi/4, 3/\sqrt{2})$ we find 8 triple points at ($\pi/2$, 4), ($0.55\pi$, 2.17), ($0.698\pi$, 1.38), ($0.715\pi$, 1.25), ($0.722\pi$, 1.01), ($1.673\pi$, 1.35), ($1.812\pi$, 2.76), and ($1.825\pi$, 2.62).

The total magnetization in field direction $\hat h = \vec h / \lvert \vec h \rvert$ is given by $\vec m \cdot \hat h/S = N^{-1} \sum_i \cos \theta_i = -N^{-1} (\partial E/\partial h)$ with $E\equiv E(\varphi,h)$ as the ground-state energy for given $\varphi$ and $h$. The magnetization curves agree very well with the MC measurements, see Fig.~\figmagXh\ in the main text. Exceptions are a few data points very close to first-order transitions; we attribute these deviations to hysteresis effects in the MC simulations.
We have explicitly checked that the minimized energy from the analytical parametrization is always less than or equal the one from the cooled MC configuration for the same parameters.

We visualize several magnetic unit cells of the spin configurations for all canted phases in Fig.~\ref{fig:spin-configs}. Fig.~\figspincfgs\ in the main text analogously shows one respective magnetic unit cell for selected phases. In Fig.~\ref{fig:spin-configs}, we display also the projections of the spin configurations onto the plane perpendicular to $\vec h$ (to be compared with the cooled MC spin configurations in Fig.~\ref{fig:mc-snapshots}).

We can also use the parametrized spin configurations to compute static spin structure factors, allowing a comparison with the MC structure factors (Fig.~\figstrfct\ in the main text). 
However, an efficient MC simulation (in our case with parallel tempering) averages over the full ground-state manifold. It consequently does not allow the direct distinction between single- and multi-$Q$ states. For example, while a pure ``$+z$ zigzag'' state with the spins of a particular zigzag line on the honeycomb lattice pointing along the $+z$ direction would exhibit a Bragg peak at only one out of the three inequivalent M points in the first Brillouin zone (M$_2$), the simulations always average over $\pm x$, $\pm y$, and $\pm z$ zigzag states (as long as these are degenerate), and the MC structure factors exhibit Bragg peaks at all three M points. 
Experimentally, this is equivalent to having multiple magnetic domains in a large sample.
Using the analytical parametrization, by contrast, we can compute ``single-domain'' structure factors for \emph{fixed} states without averaging over the ground-state manifold, allowing us to distinguish between single-$Q$ and multi-$Q$ phases in a direct way. 
In Fig.~\ref{fig:strctr-fctr_ana} we show examples for the canted zigzag phase with a Bragg peak at only one out of the three M points in the first Brillouin zone~(a), to be compared with the AF star phase which exhibits Braggs peaks at all three M points~(b). Fig.~\ref{fig:strctr-fctr_ana}(c) shows the single-domain structure factor of the zigzag star phase, with a total of 18 inequivalent Bragg peaks in the first Brillouin zone, to be compared with the MC averaged structure factor of Fig.~\figstrfct (b) in the main text.


\section{Klein duality and star vs. zigzag/stripy phases}

We explain how the fact that the AF star and zigzag states, and analogously the FM star and stripy states, are classically degenerate for all $\varphi$ can be understood in terms of the Klein duality~\cite{Cha10, Cha13, kimchi2014}. This will also allow us to gain useful insight into the quantum-fluctuation effects on the phase diagram for $S=1/2$. We introduce the dual spins $\vec{S}_i'$ by dividing the honeycomb lattice into four sublattices $\mathrm A$, $\mathrm B$, $\mathrm C$, $\mathrm D$ and identifying~\cite{Cha10}
\begin{equation} \label{eq:duality}
 \vec{S}_i' \equiv
 \begin{cases}
  \vec S_i 			& \text{for } i \in \mathrm A,\\
  \diag(1,-1,-1)\, \vec S_i 	& \text{for } i \in \mathrm B,\\
  \diag(-1,1,-1)\, \vec S_i 	& \text{for } i \in \mathrm C,\\
  \diag(-1,-1,1)\,\vec S_i 	& \text{for } i \in \mathrm D.
 \end{cases}
\end{equation}
In terms of the dual spins the Heisenberg-Kitaev Hamiltonian [Eq.~\eqhk\ in the main text] can be written as
\begin{equation} \label{eq:hk-dual}
\mathcal{H} =
- J \sum_{\left\langle ij\right\rangle}
\vec{S}_{i}' \cdot \vec{S}_{j}'
+ 2(K+J) \sum_{\left\langle ij\right\rangle_{\gamma}}
S_{i}^{\gamma}{}' S_{j}^{\gamma}{}'
- \sum_{i} \vec h_i' \cdot \vec{S}_i'.
\end{equation}
with $\vec h_i'$ denoting the dual magnetic field, obtained by a duality transformation that is analogous to the spin transformation in Eq.~\eqref{eq:duality}. Eq.~\eqref{eq:hk-dual} describes a Heisenberg-Kitaev model in a nonuniform field $\vec{h}_i'$.
For $\vec{h}_i' = 0$ and $K=-J$, i.e., $\varphi \in \{3\pi/4, 7\pi/4\}$, it features a spin $\mathrm{SU}(2)$ symmetry that is hidden in the original basis.\cite{Cha10} For finite field, a $\mathrm U(1)$ part of the hidden symmetry is left intact, if and only if $\vec h$ points along one of the cubic axes $\vec e_x$, $\vec e_y$, or $\vec e_z$ (and thus $\vec{h}_i'$ is parallel or antiparallel to this axis). For other field directions, no continuous spin symmetry remains at finite $\vec h$.

\begin{figure*}[t!]
\includegraphics[width=\textwidth]{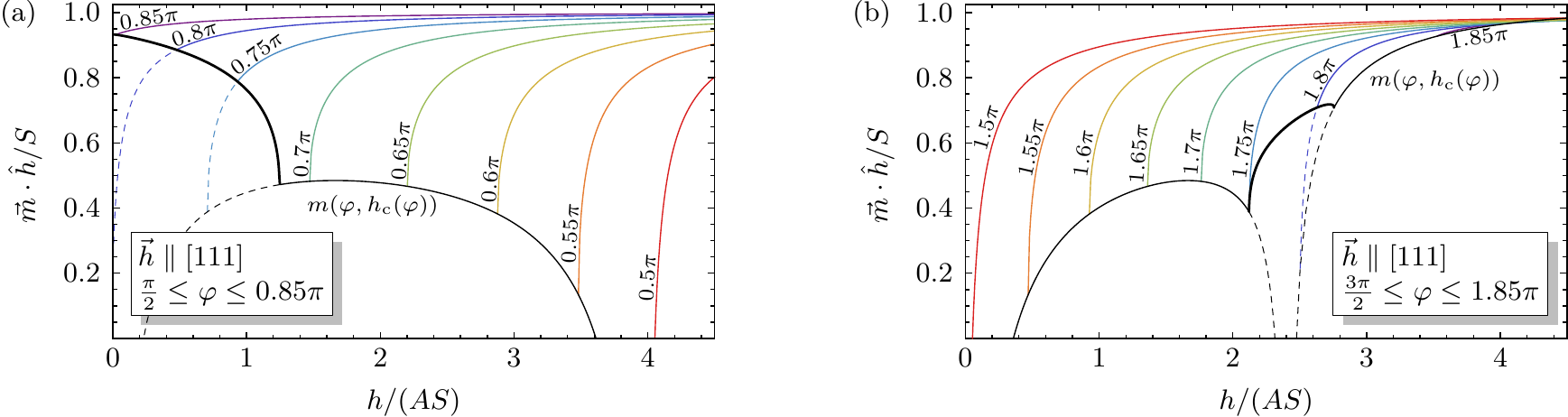}
\caption{
Total magnetization in field direction as function of applied field for $S=1/2$ in the high-field phase, $h>h_{\mathrm c}(\varphi)$, for different fixed values of $\varphi$ (a) from $\pi/2$ (red) to $0.85\pi$ (violet) above the zigzag zero-field ground state and (b) from $3\pi/2$ (red) to $1.85\pi$ (violet) above the stripy zero-field ground state. Black lines: magnetization at critical field $h_{\mathrm c}(\varphi)$. The magnetization vanishes above $h_{\mathrm c}$ for (a) $0.49\pi<\varphi<0.54\pi$ as well as for (b) $1.49\pi<\varphi<1.54\pi$, indicating strong quantum fluctuations and the breakdown of the semiclassical approximation. For $\varphi = \pi/2$ ($\varphi = 3\pi/2$) the magnetization vanishes at $h_{\mathrm{c,QSL}} \sim 4.10 AS$ ($h_{\mathrm{c,QSL}} \sim 0.053 AS$), at which one might expect a transition towards a topologically ordered spin liquid.
The dashed lines for $0.72\pi < \varphi < 0.85\pi$ and $1.75\pi < \varphi < 1.81\pi$ denote the magnetization in the metastable high-field state when the magnon instability at $\hc(\varphi)$ is preempted by a first-order transition at $h_\mathrm{c} > \hc$; here the dashed black line denotes the magnetization at $\hc(\varphi)$ and the thick black line the magnetization at $h_\mathrm{c}(\varphi)$.
}
\label{fig:magnetization-111}
\end{figure*}

Consider the exactly solvable ``stripy Klein point'' for $K=-J < 0$ (i.e., $\varphi = 7\pi/4$) and $\vec h = 0$. The quantum ground state is a ferromagnet with the dual spins pointing along a fixed, but arbitrary direction $\vec{S}_i' = S \vec n$ in spin space. Only the six states with $\vec n \in \{\pm \vec e_x, \pm \vec e_y, \pm \vec e_z\}$ out of this $\mathrm{SU}(2)$ degenerate ground-state manifold correspond to collinear spin configurations in the original basis. These are precisely the six possible stripy quantum ground states of the Heisenberg-Kitaev model. States for which only one (no) component of $\vec n$ in the cubic-axes basis vanishes correspond to coplanar (noncoplanar) spin textures in the original basis. Due to the hidden $\mathrm{SU}(2)$ symmetry an order-from-disorder mechanism can lift this quantum-ground-state degeneracy only away from the Klein point, e.g., when we consider a different set of couplings with $\varphi \notin \{3\pi/4, 7\pi/4\}$ or switch on an external field $\vec h \neq 0$.
In fact, these states belong to the highly-degenerate ground-state manifold of the classical Kitaev model,\cite{baskaran2008} and thus have the same \emph{classical} energy for all~$\varphi$.

For $7\pi/4 \leq \varphi < 1.85\pi$, we find classically that an infinitesimally small field $\vec h \parallel [111]$ lifts the degeneracy in favor of a state in which also $\vec n \parallel [111]$. In the original spin basis this state corresponds to the FM star configuration.
Upon inclusion of quantum fluctuations, one may expect that an order-from-disorder mechanism will shift the phase boundary between the stripy phase and FM star phase from zero field for $S \to \infty$ to finite values of the field for $S = 1/2$ if $\varphi > 7\pi/4$. Directly at $\varphi = 7\pi/4$, however, the degeneracy survives in the quantum case because of the presence of the hidden $\mathrm{SU}(2)$ symmetry. We infer (in the sense of degenerate perturbation theory in small $h$) that the FM star phase reaches all the way down to $h \searrow 0$, with the triple point at $(\varphi,h) = (7\pi/4,0)$ staying at zero field also for $S=1/2$.

Note that the above argument does not rely on the fact that quantum fluctuations are absent in the zero-field ground state of the dual FM model. An analogous mechanism should therefore be expected at the ``zigzag Klein point'' for $K=-J > 0$, i.e., $\varphi = 3\pi/4$. Here, the zero-field ground state in the dual basis is a N\'eel antiferromagnet with $\vec{S}_i' = (-1)^i S \vec n$ along an arbitrary direction~$\vec n$. $\vec n \in \{\pm \vec e_x, \pm \vec e_y, \pm \vec e_z\}$ corresponds to one of the six possible zigzag states in the original basis. $\vec n \parallel [111]$ corresponds to the noncoplanar AF star phase. Again, we find that a finite $\vec h \parallel [111]$ lifts the degeneracy in favor of the state with
$\vec n \parallel [111]$. Due to the absence of an order-from-disorder mechanism in the hidden-$\mathrm{SU}(2)$-symmetric model when $\varphi = 3\pi/4$ and $\vec h=0$, we expect that the degenerate zero-field ground state gives way to an AF star ground state at infinitesimal field in the $[111]$ direction also in the quantum limit when $S=1/2$.

We conclude that finite regions of both FM star and AF star phases exist for field in the $[111]$ direction not only classically, but also in the quantum phase diagram for $S=1/2$, at least in the vicinity of the Klein points at $\varphi = 3\pi/4$ or $7\pi/4$.%
\cite{noteduality}


\section{High-field magnetization for $S=1/2$}
\begin{figure*}[!t]
\includegraphics[width=\textwidth]{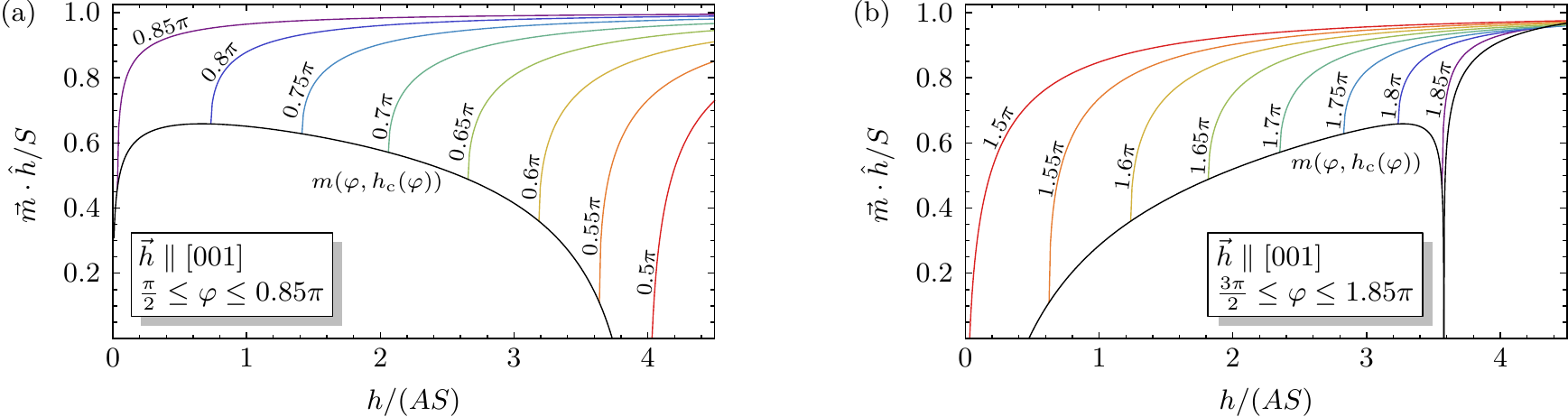}
\caption{Same as Fig.~\ref{fig:magnetization-111} for $\vec h \parallel [001]$. For $\varphi = \pi/2$ ($\varphi = 3\pi/2$) the magnetization now vanishes at $h_{\mathrm{c,QSL}} \sim 4.03 AS$ ($h_{\mathrm{c,QSL}} \sim 0.032 AS$). It also vanishes for $\varphi =  1.852 \pi$ and $h = h_\mathrm{c} = 3.58 AS$, when the instability wavevector of the high-field magnon spectrum changes between $\vec Q = \Gamma$ and $\vec Q = \mathrm M$. Here, no metastable states exist since the transition from the polarized phase towards a canted phase is always continuous for any $\varphi$.
}
\label{fig:magnetization-001}
\end{figure*}

In the polarized phase, the influence of quantum fluctuations for $S=1/2$ may be estimated by again employing spin-wave theory. To this end, we compute the $1/S$ correction to the total magnetization in field direction:
\begin{align}
\frac{\vec m \cdot \hat h}{S} & = \frac{1}{N} \left[\sum_{i \in \mathrm A}\left(1-\frac{1}{S} \langle a_i^\dagger a_i \rangle \right) + \sum_{j \in \mathrm B}\left(1-\frac{1}{S} \langle b_j^\dagger b_j \rangle \right) \right] \nonumber \\ 
 & \quad + \mathcal O(1/S^2),
\end{align}
where $\langle a_i^\dagger a_i \rangle$ and $\langle b_j^\dagger b_j \rangle$ are the magnon densities at lattice sites $i \in \mathrm A$ and $j \in \mathrm B$, respectively. The magnetization curves in linear spin-wave theory for $S=1/2$ as function of $h>h_{\mathrm c}$ in the polarized phase are depicted for selected values of $\varphi$ and $\vec h \parallel [111]$ in Fig.~\figmagXh\ in the main text. Analogous curves for full ranges of $\varphi$ in the polarized phase above the zigzag and stripy zero-field ground states are given in Fig.~\ref{fig:magnetization-111} for $\vec h \parallel [111]$ and Fig.~\ref{fig:magnetization-001} for $\vec h \parallel [001]$. In the nonfrustrated FM and AF Heisenberg cases the corrections vanish, $\langle a_i^\dagger a_i \rangle = \langle b_j^\dagger b_j \rangle = 0$, but they become enhanced by increasing the Kitaev exchange $K$.

Consider the limit $|K|\gg|J|$: Here the zero-field state for $S=1/2$ is a gapless $\Ztwo$ spin liquid, and it is known \cite{kitaev06, jiang2011} that this state is unstable towards a gapped topologically ordered spin liquid for infinitesimal field in the $[111]$ direction. Increasing $h$ eventually drives a transition towards the polarized phase at some finite $h = h_\mathrm{c,QSL} > 0$. While a quantitative analysis of this topological quantum transition is beyond the realm of linear spin-wave theory, a simple estimate for the transition points may be obtained by computing the parameter sets $(\varphi, h)$ at which the magnetization (to first order in $1/S$) vanishes. This way, we find, e.g., for $\varphi = 3\pi/2$ (FM Kitaev model) the critical field strength as $h_\mathrm{c,QSL}/(AS) \sim 0.053$, which is in about $30\%$ agreement with the value from density-matrix renormalization group calculations.\cite{jiang2011} For $\varphi = \pi/2$ (AF Kitaev model) we find a significantly higher estimate of $h_\mathrm{c,QSL} / (AS) \sim 4.10$, i.e., the spin liquid in the AF Kitaev model is much more stable against uniform applied field as compared to the FM Kitaev model.

For intermediate $|K| \sim |J|$ (e.g., near the ``Klein'' points at $\varphi = 3\pi/4$ and $7\pi/4$) the magnetization in the polarized phase is finite for all $h > h_\mathrm{c}(\varphi)$, with the leading-order correction to the saturated magnetization of the order of $50\%$. (The exception is $\varphi = 1.852\pi$ for field in the $[001]$ direction, when the instability wavevector changes from $\Gamma$ to M, the lower magnon band becomes flat with $\varepsilon_{\vec q} = 0$ between $\Gamma$ and M for $h \searrow h_\mathrm{c}$, and the leading-order magnetization correction diverges.)
For the case of possible experimental relevance, $\varphi \sim (0.6\dots0.72) \pi$, we hence expect that the gapped high-field phase is reached at a magnetization of about half the saturation magnetization.
